\documentclass[aps,amsmath,floats,floatfix, twocolumn,
superscriptaddress,nofootinbib,showpacs]{revtex4-1}

\usepackage[colorlinks, pdfborder={0 0 0}, plainpages=false, citecolor=blue]{hyperref}
\usepackage{graphicx}
\usepackage{xspace}
\usepackage[usenames,dvipsnames]{color}
\usepackage{amssymb}
\usepackage[normalem]{ulem} %for \sout

\newcommand{\Caltech}{\affiliation{Theoretical Astrophysics 350-17,
    California Institute of Technology, Pasadena, CA 91125, USA}}
\newcommand{\CITA}{\affiliation{Canadian Institute for Theoretical
    Astrophysics, 60 St.~George Street, University of Toronto,
    Toronto, ON M5S 3H8, Canada}} %

\newcommand{\Cornell}{\affiliation{Cornell Center for Astrophysics
    and Planetary Science, Cornell University, Ithaca, NY 14853, USA}}
\newcommand{\UmassD}{\affiliation{Mathematics Department, University of
  Massachusetts Dartmouth, Dartmouth, MA 02747, USA}}
\newcommand{\Kyoto}{\affiliation{Center for Gravitational Physics
and International Research Unit of Advanced Future Studies,
Yukawa Institute for Theoretical Physics, Kyoto University, Kyoto, Japan}}
\newcommand{\JPL}{\affiliation{Caltech JPL, Pasadena, California 91109, USA}}

\begin{document}

\title{
A Numerical Relativity Waveform Surrogate Model for Generically Precessing
Binary Black Hole Mergers
}

\author{Jonathan Blackman} \Caltech
\author{Scott E. Field} \UmassD \Cornell
\author{Mark A. Scheel} \Caltech
\author{Chad R. Galley} \Caltech
\author{Christian D. Ott} \Caltech \Kyoto
\author{Michael Boyle}\Cornell
\author{Lawrence~E.~Kidder}\Cornell
\author{Harald P.~Pfeiffer}\CITA
\author{B\'{e}la Szil\'{a}gyi}\Caltech\JPL

\date{\today}

\begin{abstract}
A generic, non-eccentric binary black hole (BBH) system emits gravitational waves (GWs)
that are completely described by 7 intrinsic parameters:
the black hole spin vectors and the ratio of
their masses.
Simulating a BBH coalescence by solving Einstein's equations numerically is
computationally expensive, requiring days to months of computing
resources for a single set of parameter values.
Since theoretical predictions of the GWs are often needed for
%millions of
many different source parameters, a fast and accurate model is essential.
We present the first surrogate model for GWs from the
coalescence of BBHs including all
$7$ dimensions of the intrinsic non-eccentric parameter space.
The surrogate model, which we call NRSur7dq2, is built from the results of $744$
numerical relativity simulations.
NRSur7dq2 covers spin magnitudes up to $0.8$ and mass ratios up to $2$,
includes all $\ell \leq 4$ modes, begins about $20$ orbits
before merger, and can be evaluated in $\sim~50\,\mathrm{ms}$.
We find the largest NRSur7dq2 errors to be comparable to the largest errors
in the numerical relativity simulations, and more than an order of
magnitude smaller than the errors of other waveform models.
Our model, and more broadly the 
methods developed here, will enable studies that were not previously
possible when using highly accurate
waveforms, such as parameter inference
and tests of general relativity with GW observations.
\end{abstract}

% 04.25.D- Numerical relativity
% 04.25.dg Numerical studies of black holes and black-hole binaries
% 04.25.Nx Post-Newtonian approximation; perturbation theory; related approximations 
% 04.30.-w Gravitational waves (see also 04.80.Nn Gravitational wave detectors and experiments)
% 04.30.Db Wave generation and sources 
% 02.70.Hm Spectral methods

\maketitle

%%%%%%%%%%%%%%%%%%%%%%%%%%%%%%%%%%%%%%%%%%%%%%%%%%%%%%%%%%%%%%%%%%%%%%%%%%%%%%%
\section{Introduction}
%%%%%%%%%%%%%%%%%%%%%%%%%%%%%%%%%%%%%%%%%%%%%%%%%%%%%%%%%%%%%%%%%%%%%%%%%%%%%%%

With LIGO's two confident detections of gravitational waves (GWs)
from binary black hole (BBH) systems \cite{LIGOVirgo2016a,Abbott:2016nmj},
we have entered the exciting new era of GW astronomy.
The source black hole (BH) masses and spins can be
determined by comparing the signal to waveforms
predicted by general relativity (GR) ~\cite{TheLIGOScientific:2016wfe, Abbott:2016apu},
and new strong-field tests of GR can be performed~\cite{TheLIGOScientific:2016src}.
These measurements and tests
require GW models that are both accurate and fast to evaluate.
The total mass of the system $M$ can be scaled out of the problem,
leaving a $7$-dimensional non-eccentric intrinsic
parameter space over which the waveform
must be modeled, consisting of the mass ratio and two BH spin vectors.

Numerical relativity (NR) simulations of BBH
mergers~\cite{Pretorius2005a, Zlochower:2005bj, SpECwebsite,
einsteintoolkit, Husa2007, Bruegmann2006, Herrmann2007b}
solve the full Einstein equations and produce the most accurate waveforms.
These simulations are computationally expensive, requiring weeks to months
on dozens of CPU cores
for a waveform beginning $\sim 20$ orbits before the merger.
Analytic and semi-analytic waveform
models~\cite{Damour:024009, Damour2009a, Taracchini:2013rva, Purrer:2015tud,
             Pan:2013rra, Bohe:2016gbl, Hannam:2013oca, Khan:2015jqa,
             Husa:2015iqa}
are quick to evaluate, but they make approximations
  that can introduce differences with respect to the true waveform
  predicted by GR.
These differences
could lead to parameter biases or inaccurate tests of GR for some
high signal-to-noise ratio detections that could be made in the near
future~\cite{Lindblom2008, Abbott:2016wiq}.

A {\em surrogate waveform model}~\cite{Blackman:2017dfb,
    Blackman:2015pia,Field:2013cfa,
    Purrer:2014,Purrer:2015tud}
is a model that takes a set of
precomputed waveforms that were generated by some other model (e.g., NR
or a semianalytic model), and interpolates in parameter space between
these waveforms to quickly produce a waveform for any desired
parameter values.  A surrogate waveform can be evaluated much more quickly
than the underlying model, and can be made as accurate as the
underlying model given a sufficiently large set of precomputed
waveforms that cover the parameter space.
Previous surrogate models based on
NR waveforms were built for non-spinning BBH
systems~\cite{Blackman:2015pia} and for a
$4$-dimensional ($4d$) parameter subspace containing
precession~\cite{Blackman:2017dfb}.
Here, we present the first
NR surrogate model including all $7$
dimensions of the parameter space.
The model, which we call NRSur7dq2, produces waveforms nearly as accurate
as those from NR simulations, but can be evaluated in $\sim~50\,\mathrm{ms}$ on
a single CPU core for a speedup of more than $8$ orders of magnitude
compared to NR.
Our method enables performing high accuracy GW data analysis,
including parameter inference for astrophysics and tests of GR.

%%%%%%%%%%%%%%%%%%%%%%%%%%%%%%%%%%%%%%%%%%%%%%%%%%%%%%%%%%%%%%%%%%%%%%%%%%%%%%%
\section{Numerical Relativity Data}
%%%%%%%%%%%%%%%%%%%%%%%%%%%%%%%%%%%%%%%%%%%%%%%%%%%%%%%%%%%%%%%%%%%%%%%%%%%%%%%

The NR simulations used to build the surrogate model are performed using
the Spectral Einstein Code
(SpEC)~\cite{SpECwebsite,
Pfeiffer2003, Lovelace2008, Lindblom2006, Szilagyi:2009qz, Scheel2009,
Szilagyi:2014fna}.
The simulations begin at a coordinate time $\tau=0$, where we specify
the BH mass ratio $q = m_1/m_2 \geq 1$
and initial dimensionless spin vectors
\begin{equation}
\vec{\chi}_i(\tau=0) = \vec{S}_i(\tau=0) / m_i^2\,,\,\, i \in \{1, 2\}\,.
\end{equation}
%\Note{JB: From here to the end of this paragraph is all identical to
%the 4d2s case, except here we use the center of mass correction. Can
%we remove all these details and say it's all the same as 4d2s except
%we also remove the lienar drifts?}
The system is evolved through merger and ringdown, and the GWs
are extracted at multiple finite radii from the source.
These are extrapolated to future null infinity~\cite{Boyle-Mroue:2008}
using quadratic polynomials in $1/r$, where $r$ is a radial coordinate.
The effects of any drifts in the center of mass that are linear in time
are removed from the
waveform~\cite{Boyle2015a,Boyle:2013a,Boyle:2014,scri}.
The waveforms at future null infinity use a time coordinate $\tilde{\tau}$,
which is different from the simulation time $\tau$, and begins approximately at
$\tilde{\tau}=0$.
The spins $\vec{\chi}_i(\tau)$ are also measured at each simulation time.
To compare spin and waveform features, we identify $\tau$ with $\tilde{\tau}$.
While this identification is not gauge-independent, the spin directions
are already gauge-dependent.
We note that the spin and orbital angular momentum vectors in the damped
harmonic gauge used by SpEC agree quite well with the corresponding vectors
in post-Newtonian (PN) theory~\cite{Ossokine:2015vda}.

Once we have the spins $\vec{\chi}_i(\tau)$ and spin-weighted spherical
harmonic modes of the waveform $h^{\ell, m}(\tau)$,
we perform the same alignment discussed in Sec.~III.D of
Ref.~\cite{Blackman:2017dfb}.
Briefly, 
for each simulation, 
we first determine the time $\tau_\mathrm{peak}$ which maximizes the
total amplitude of the waveform
\begin{equation}
A_\mathrm{tot}(\tau) = \sqrt{\sum_{\ell, m}|h^{\ell, m}(\tau)|^2}\,.
\end{equation}
We determine $\tau_\mathrm{peak}$ by fitting a quadratic function
to $5$ adjacent samples of $A_\mathrm{tot}(\tau)$, consisting of the largest sample and two neighbors on either side.
We choose a new time coordinate
\begin{equation}
t = \tau - \tau_\mathrm{peak}\,,
\end{equation}
which maximizes $A_\mathrm{tot}$ at $t=0$.
We then rotate the waveform modes such that at our reference time
of $t=t_0=-4500M$, $\hat{z}$ is the principal
eigenvector of the angular momentum operator~\cite{Boyle:2011gg}
and the phases of $h^{2, 2}(t_0)$ and $h^{2, -2}(t_0)$ are equal.
We sample the waveform and spins in steps of $\delta t=0.1M$,
from $t_0$ to $t_f=100M$, by interpolating the 
real and imaginary parts of each waveform mode, 
as well as the spin components, using cubic splines.
The initial separations and velocities of the BBH systems
were chosen such that, after aligning the peak amplitude to $t=0$
as above, the waveforms begin at $t \approx -5000M$.
Choosing $t_0=-4500M$ discards the first $\sim 500M$ which is
contaminated by junk radiation~\cite{Aylott:2009ya}.

We first include
all $276$ NR simulations used in the NRSur4d2s surrogate model
and the $9$ additional simulations used in Sec.~IV.D and Table V of
Ref.~\cite{Blackman:2017dfb}.
We perform $459$ additional NR simulations.
The first $361$ of these are chosen based on sparse
grids~\cite{smolyak1963quadrature, bungartz2004sparse} and include combinations of
extremal parameter values (such as $q \in \{1, 2\}$) and intermediate
values as detailed in Appendix~\ref{app:parameters}.
The parameters for the remaining $98$ simulations are chosen
as follows.
We randomly sample $1000$ points in parameter space
uniformly in mass ratio,
spin magnitude, and spin direction on the sphere.
We compute the distance between points $a$ and $b$ using
\begin{equation}
ds^2 = \left(0.3\left(q_a - q_b\right)\right)^2
    + \sum_{i\in\{1, 2\}}\left\|\vec{\chi}_{ia} - \vec{\chi}_{ib}\right\|^2\,.
\end{equation}
The coefficients multiplying each term in this expression have been 
chosen somewhat arbitrarily, although our expectation is that any choice 
of order unity should provide a reasonable criteria for point selection.
For each sampled parameter, we compute the minimum distance to all
previously chosen parameters.
We then choose the sampled parameter maximizing this minimum distance.
We then resample the $1000$ parameters for the next of the $98$ iterations.
This results in a total of $744$ NR simulations.
For simulations with equal masses and unequal spins, we use the results twice
by reversing the labeling of the BHs and rotating the waveform
accordingly.
There are $142$ such simulations, leading to $886$ NR waveforms.

%%%%%%%%%%%%%%%%%%%%%%%%%%%%%%%%%%%%%%%%%%%%%%%%%%%%%%%%%%%%%%%%%%%%%%%%%%%%%%%
\section{Waveform Decomposition}
%%%%%%%%%%%%%%%%%%%%%%%%%%%%%%%%%%%%%%%%%%%%%%%%%%%%%%%%%%%%%%%%%%%%%%%%%%%%%%%

The goal of a surrogate model is to
  take a precomputed set of
waveform modes $\{h_i^{\ell, m}(t)\}$ at a fixed set of
  points in parameter space $\{\vec{\lambda}_i\}$,
and to produce waveform modes $\{h^{\ell, m}(t)\}$
  at new desired parameter values.  Because
  $h^{\ell, m}(t)$ is highly oscillatory and changes
  in a complicated way as one varies the masses and spins, it is not
  feasible to directly interpolate $\{h_i^{\ell, m}(t)\}$
in parameter space
  with only $886^{1/7} \approx 2.64$ available points per dimension.
  Instead, we decompose each waveform $h(t)$ into many
  \emph{waveform data pieces}. Each waveform data piece is a simpler
  function that varies slowly over parameters. Once we have
  interpolated each waveform data piece to a desired point in parameter
  space, we recombine them to form $h(t)$.  Our
  decomposition is similar to but improves upon the one used in
  Ref.~\cite{Blackman:2017dfb}.

%\jon{The waveform modes $h^{\ell, m}(t)$ oscillate in time, as seen in
%Fig.~\ref{fig:waveforms}. If we choose a fixed time $t$ and vary the
%mass ratio and spins, the modes also oscillate rapidly over the $7d$ parameter
%space.
%With only $886^{1/7} \approx 2.64$ available points per dimension,
%there is no hope of fitting the waveform modes directly over the parameter
%space.
%\sout{
%The NRSur4d2s surrogate used}
%We instead use} a \emph{waveform decomposition} that
%\jon{\sout{transformed} transforms}
%the oscillatory waveform modes $\{h^{\ell, m}(t)\}$ to many more slowly varying
%\emph{waveform data pieces}\jon{\sout{ which were modeled directly}}.
%\jon{Our decomposition is similar to but improves upon
%the one used in Ref.~\cite{Blackman:2017dfb}.
%\sout{Here we make similar but different transformations.}}
  We first determine the unit quaternions $\hat{q}(t)$ that
    define the coprecessing frame~\cite{Schmidt2010, OShaughnessy2011, Boyle:2011gg},
and we determine the waveform modes $\{h_C^{\ell, m}(t)\}$ in this frame.
This is done using the transformation $T_C$
given by Eq.~25 of Ref.~\cite{Blackman:2017dfb}.
%We do, however, filter the
%  spin components $\vec{\chi}_i(t)$, using the same Gaussian
%low-pass filter described in Eqs. 41--47 of
%Ref~\cite{Blackman:2017dfb}. \Mark{The purpose of the filtering is ...}
%The effect of filtering can be seen by comparing the thick blue line (unfiltered)
%to the dashed orange line (filtered) in Fig.~\ref{fig:spins}.
%\jon{\sout{The spins $\vec{\chi}_i(t)$ are also transformed
%to the coprecessing frame using}}
%\begin{equation}
%\jon{\sout{\vec{\chi}_i^\mathrm{copr}(t) = \hat{q}^{-1}(t) \vec{\chi}_i(t)
%                                \hat{q}(t)\,.}}
%\end{equation}
%\jon{\sout{
%Note that quaternion multiplication is used here, and vectors are treated
%as quaternions with zero scalar component.
%We note that unlike in Ref.~\cite{Blackman:2017dfb},
%here we do not filter $\hat{q}(t)$ and, as described below, 
%the spins play an important role in the surrogate's construction.
%}}
%\Note{JB: There used to be a paragraph break here. I removed it
%    due to the strikeouts above.}
The orbital phase
\begin{equation}
\label{eq:orbphase}
\varphi(t) = \frac{1}{4}\left(\arg\left[h_C^{2, -2}(t)\right] -
                              \arg\left[h_C^{2, 2}(t)\right]\right) \,,
\end{equation}
is computed from the coprecessing waveform modes.
This is expected to be
superior to computing the orbital phase from
the BH trajectories because unlike the
coordinate-dependent trajectories, the waveform can be made
gauge invariant up to Bondi-Metzner-Sachs transformations~\cite{Boyle2015a}.
%\Note{JB: This paragraph break is new}

We filter the spins $\vec{\chi}_i$ in the inertial frame using the same Gaussian
filter that was used to filter $\hat{q}(t)$ in Ref.~\cite{Blackman:2017dfb},
and note that here we do not filter $\hat{q}(t)$.
The filtered spins are given by
\begin{equation}
\vec{\chi}_i^\mathrm{filt}(t) = \frac{\int_{\varphi_-}^{\varphi_+}
    \vec{\chi}_i(t'(\varphi)) G(\varphi) d\varphi}{
    \int_{\varphi_-}^{\varphi_+}G(\varphi)d\varphi}\,,
\end{equation}
where $t'(\varphi)$ is the inverse function of Eq.~\ref{eq:orbphase},
and $\varphi_\pm$ and $G(\varphi)$ are given by Eqs. 39-43 of
Ref.~\cite{Blackman:2017dfb}.
Note that $\varphi_\pm$ and $G(\varphi)$ implicitly depend on $t$.
The spins are then transformed
to the coprecessing frame using
\begin{equation}
\vec{\chi}_i^\mathrm{copr}(t) = \hat{q}^{-1}(t) \vec{\chi}_i^\mathrm{filt}(t)
                                \hat{q}(t)\,.
\end{equation}
Note that quaternion multiplication is used here, and vectors are treated
as quaternions with zero scalar component.
We find that filtering the spins leads to a more accurate surrogate model
by suppressing the orbital timescale oscillations in the spin
components~\cite{Ossokine:2015vda} and therefore making the spin time derivatives
easier to model.

We
then transform the spins and waveform modes to a coorbital frame, in
which the BHs are nearly on the $x$ axis.
The coorbital frame is just the coprecessing frame rotated by $\varphi(t)$ about
the $z$ axis.
Specifically, we have
\begin{align}
\hat{q}_\mathrm{r}(t) &= \cos\left(\frac{\varphi(t)}{2}\right) +
                      \hat{z}\sin\left(\frac{\varphi(t)}{2}\right)\,, \\
\vec{\chi}_i^{\mathrm{coorb}}(t) &= \hat{q}^{-1}_\mathrm{r}(t)
                                    \vec{\chi}_i^\mathrm{copr}(t)
                                    \hat{q}_\mathrm{r}(t)\,, \\
\label{eq:coorb_spins}
h^{\ell, m}_{\mathrm{coorb}}(t) &= h^{\ell, m}_C(t)e^{im\varphi(t)}\,,
\end{align}
where $\hat{q}_\mathrm{r}(t)$ is a unit quaternion representing a rotation
about the $\hat{z}$ axis by $\varphi$.
Finally, using $4$th order finite differences, we compute the orbital frequency
\begin{equation}
  \label{eq:defomega}
\omega(t) = \frac{d}{dt}\varphi(t)
\end{equation}
and the spin time derivatives in the coprecessing frame, which we then transform
to the coorbital frame
\begin{equation}
%\dot{\vec{\chi}}_i(t) &= \frac{d}{dt} \vec{\chi}_i^\mathrm{copr}(t)\,,\\
\dot{\vec{\chi}}_i^{\mathrm{coorb}}(t) = \hat{q}^{-1}_\mathrm{r}(t)
                                    \dot{\vec{\chi}}_i^\mathrm{copr}(t)
                                    \hat{q}_\mathrm{r}(t)\,,
\label{eq:spin_rotation}
\end{equation}
where a dot means $d/dt$.
For the precession dynamics, we compute
the angular velocity of the coprecessing frame
\begin{align}
\frac{1}{2}\vec{\Omega}^\mathrm{copr}(t) &= \lim_{dt \rightarrow 0} \frac{1}{dt}
        \left(\hat{q}^{-1}(t) \hat{q}(t + dt) - 1\right)  \\
                &= s(t)\dot{\vec{v}}(t) - \dot{s}(t)\vec{v}(t) -
                    \vec{v}(t)\times\dot{\vec{v}}(t) \label{eq:13} \,,
\end{align}
where $s(t)$ and $\vec{v}(t)$ are the scalar and vector components of $\hat{q}(t)$.
Fourth-order 
finite difference stencils are used to compute the time derivatives 
appearing in Eq.~\eqref{eq:13}.
We also transform $\vec{\Omega}^\mathrm{copr}(t)$ to the coorbital frame to obtain
$\vec{\Omega}^\mathrm{coorb}(t)$ as in Eq.~(\ref{eq:spin_rotation}).
The minimal rotation condition of the coprecessing frame ensures
\begin{equation}
\Omega_z^\mathrm{coorb}(t) = \Omega_z^\mathrm{copr}(t) = 0
\end{equation}
up to finite difference errors.

\begin{figure}
    \includegraphics[width=\linewidth]{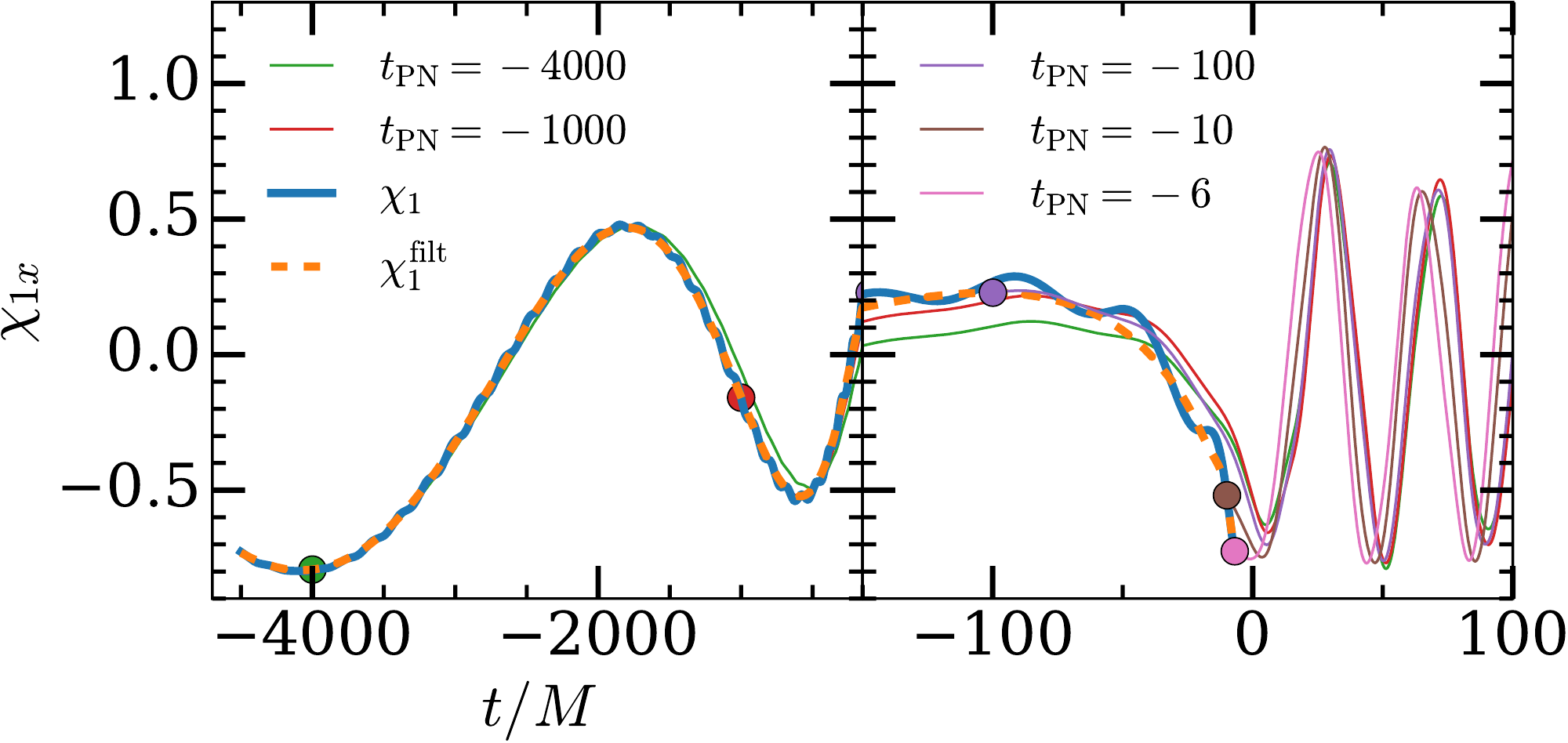}
    \caption{
        The $x$-component of a spin extended through merger and
        ringdown with PN expressions.
        These spins are not physically meaningful, but provide
        a parameterization of the system leading to accurate fits.
        The thick solid blue curve and dashed orange curve
            show the unfiltered and filtered spins
        from an NR simulation, and are
        not measured past $t=-6M$ due to the merger of the BHs.
        Each thin line is identical to the filtered NR curve before some
        time $t_\mathrm{PN}$ indicated with a dot, after which the spins are
        evolved using Eq.~(\ref{eq:spins}).
        The spins during the ringdown are affected somewhat by the choice
        of $t_\mathrm{PN}$, but the overall phasing is quite similar.
    }
    \label{fig:spins}
\end{figure}

Given a waveform data piece $X(t)$ evaluated at a set of parameters,
one would be tempted to parameterize $X(t)$ at any fixed time $t_i$
by the mass ratio and the initial spins, and then construct a fit to
$X(t_i)$ as a function of these parameters.  However, we find
much better fits if we instead parameterize $X(t_i)$ by
the spins \emph{at time $t_i$} and the mass ratio.  While this is easy to
do during the inspiral where we still have two BHs with
individual spins, we seek a way to extend this parameterization
through the merger and ringdown, where individual BH spins
are no longer available.
We extend, unphysically,
  the spin evolution through the merger and ringdown using the PN
expressions
\begin{equation}
\frac{d}{dt}\vec{\chi}_{i} = \vec{\Omega}_i^\mathrm{Spin} \times \vec{\chi}_i\,,
\label{eq:spins}
\end{equation}
where $\vec{\chi}_{i}$ is the spin in the inertial frame, and
$\vec{\Omega}_i^\mathrm{Spin}$ is a PN expression
given by Eq.~A32 of
Ref.~\cite{Ossokine:2015vda}. $\vec{\Omega}_i^\mathrm{Spin}$
is a function of
the orbital angular momentum vector $\hat{l}(t)$,
a vector pointing from one BH to the other $\hat{n}(t)$, and the
PN parameter $x(t)$.
Evaluating $\vec{\Omega}_i^\mathrm{Spin}$
requires several quantities that are typically
computed from BH trajectories in PN theory.
Since the trajectories are also not available after the merger, we compute
them from the waveform.
We take
$\hat{l}$ and $\hat{n}$
to be the $\hat{z}$ and $\hat{x}$ axes of the coorbital frame, and we
take the PN parameter $x$ to be $\omega^{2/3}$,
where $\omega$ is defined in Eq.~(\ref{eq:defomega})
(see Eq. 230 of Ref.~\cite{Blanchet2014}).
We choose $t_\mathrm{PN}=-100M$ and begin the PN integrations from the
spins at $t_\mathrm{PN}$.
The extended spins are somewhat robust to the choice of $t_\mathrm{PN}$ as
seen in Fig.~\ref{fig:spins}.
We stress that these extended spins are not physically meaningful
for $t > t_\mathrm{PN}$, but
provide a convenient parameterization of the system that leads to
accurate parametric fits.

%%%%%%%%%%%%%%%%%%%%%%%%%%%%%%%%%%%%%%%%%%%%%%%%%%%%%%%%%%%%%%%%%%%%%%%%%%%%%%%
\section{Building the Model}
%%%%%%%%%%%%%%%%%%%%%%%%%%%%%%%%%%%%%%%%%%%%%%%%%%%%%%%%%%%%%%%%%%%%%%%%%%%%%%%

In this section, we describe the quantities that are computed from
the waveform data pieces and
stored when building the NRSur7dq2 surrogate model.
The subsequent section will then describe how the NRSur7dq2 surrogate
model uses these stored quantities to generate waveforms.

We first construct
surrogate models for the waveform modes
in the coorbital frame $h^{\ell, m}_\mathrm{coorb}(t)$.
For $m=0$ modes, we directly 
model the real and imaginary components without any additional
decompositions.
For $m > 0$, we compute
\begin{equation}
h^{\ell, m}_\pm = \frac{1}{2}\left(h^{\ell, m}_\mathrm{coorb} \pm
        h^{\ell, -m\,*}_\mathrm{coorb}\right)
\end{equation}
and model the real and imaginary parts of $h^{\ell, m}_\pm$.
Each of these modeled components is considered a \emph{waveform data piece}.
We proceed according to Sec.~V of Ref.~\cite{Blackman:2017dfb}:
For each waveform data piece,
we construct a compact linear basis using singular
value decomposition with a RMS tolerance of $3\times 10^{-4}.$
We then construct an empirical interpolant and determine one empirical node time
$T_j$ for each basis vector.
The times $T_j$ are
chosen differently for each waveform data piece.
Finally, for each $T_j$,
we construct a \emph{parametric fit}
for the waveform data piece evaluated
at $T_j$, which is described below.
The fits are functions of the mass ratio and the coorbital spin components
$\vec{\chi}^\mathrm{coorb}_i(t)$ evaluated at $T_j$.
Note that the $x$ component of a vector in the coorbital frame
is roughly the component in the direction of a
vector pointing from one BH to the other, the $z$ component is along
the axis of orbital angular momentum, and the $y$ component is the remaining
orthogonal direction.
In addition to the resulting fit data, the empirical interpolation matrix
(see Eq.~(B7) of Ref.\cite{Field:2013cfa})
for each of these waveform data pieces is
stored in the NRSur7dq2 surrogate model.

These parametric fits 
use the forward-stepwise greedy fitting method described in Appendix A of
Ref.~\cite{Blackman:2017dfb}.
One benefit of this fitting method is 
that it automatically selects 
for higher order fits whenever the data 
is high-quality (e.g. in the inspiral), and lower order fits
whenever the data is more noisy (e.g. during the ringdown).
We choose the basis functions %$B_l^k(\lambda^l) = (\lambda^l)^k$
to be a tensor product of 1D monomials
in the spin components and
\begin{equation}
x = \frac{q - 1.5}{0.51}\,,
\end{equation}
which is an affine mapping from $[0.99, 2.01]$
to the standard interval $[-1,1]$.
%where $\lambda$ is a $7d$ parameter vector indexed by $l$
%containing the mass ratio and spin components.
We consider up to cubic functions in $x$
and up to quadratic
functions in the spin components.
We perform $20$ trials using $50$ validation points each.
The fit coefficients and the basis functions selected
during the fitting procedure
are stored in the NRSur7dq2 surrogate model.

{We also construct parametric fits for $\omega(t)$,
$\Omega_{\{x, y\}}^\mathrm{coorb}(t)$, and
$\dot{\chi}_{j\{x, y, z\}}^\mathrm{coorb}(t)$ at selected time nodes $t_i$.
These quantities describe
the dynamics of the binary and the spins, so we call these
$t_i$ the dynamics time nodes.
We attempt to choose the time
nodes $t_i$ to be approximately uniformly spaced in $\varphi(t)$
with $10$ nodes per orbit.
Because $\varphi(t)$ is different for different simulations, and
we choose the same time nodes for all simulations, in practice
our choice of $238$ time nodes gives us between $8$ and $15$ nodes per orbit.
We find that this is sufficient ---
including additional nodes per orbit does not improve the accuracy of
the surrogate model.  Our time nodes are labeled
$t_0 < t_1 < \dots < t_{234}=100M$ plus
three additional nodes $t_\frac{1}{2}$, $t_\frac{3}{2}$, and $t_\frac{5}{2}$,
which are the midpoints of their adjacent integer time nodes.  The
reason for including the fractional time nodes is for Runge-Kutta
time integration at the beginning of the time series,
which will be made clear in the next section.
In Appendix~\ref{app:timesamples},
we describe in detail the algorithm for  choosing $t_i$, but any choice
  that is roughly uniformly-spaced in $\varphi(t)$ and sufficiently dense
  should yield a surrogate with comparable accuracy.

%%%%%%%%%%%%%%%%%%%%%%%%%%%%%%%%%%%%%%%%%%%%%%%%%%%%%%%%%%%%%%%%%%%%%%%%%%%%%%%
\section{Evaluating the model}
%%%%%%%%%%%%%%%%%%%%%%%%%%%%%%%%%%%%%%%%%%%%%%%%%%%%%%%%%%%%%%%%%%%%%%%%%%%%%%%
\label{sec:evaluating-model}

To evaluate the NRSur7dq2 surrogate model, we provide the mass ratio $q$
and initial spins $\vec{\chi}_j(t_0)$ as inputs.
The evaluation consists of three steps: we first integrate
a coupled ODE system for the spins, the orbital phase, and the
coprecessing frame, then we evaluate the coorbital waveform modes,
and finally we transform
the waveform back to the inertial frame.
We describe each of these steps below.

We initialize the ODE system with
\begin{align*}
\varphi(t_0)=0 \,, \quad \hat{q}(t_0) = 1 \,, \quad
\vec{\chi}_j^\mathrm{copr}(t_0) = \vec{\chi}_j(t_0) \,.
\end{align*}
%\begin{itemize}
%  \item $\varphi(t_0)=0$ ,
%  \item $\hat{q}(t_0) = 1$ ,
%  \item $\vec{\chi}_j^\mathrm{copr}(t_0) = \vec{\chi}_j(t_0)$ .
%\end{itemize}
To integrate this system forward in time using a numerical
ODE solver (described below), we need to evaluate the
time derivatives of $\varphi$, $\hat{q}$, and $\vec{\chi}_j^\mathrm{copr}$
at a time node $t_i$, given the values of those variables at $t_i$.
To do this,
we first determine $\vec{\chi}_j^\mathrm{coorb}(t_i)$
by rotating the $x$ and $y$ components of $\vec{\chi}_j^\mathrm{copr}(t_i)$
by an angle $\varphi(t_i)$ as in Eq.~(\ref{eq:coorb_spins}).
We then evaluate the fits for $\omega(t_i)$,
$\Omega^\mathrm{coorb}_{\{x, y\}}(t_i)$, and
$\dot{\chi}_{j\{x, y, z\}}^\mathrm{coorb}(t_i)$
using the mass ratio $q$ and the current coorbital spins
$\vec{\chi}_j^\mathrm{coorb}(t_i)$.
We set $\Omega^\mathrm{coorb}_z(t_i)=0$,
and obtain $\dot{\vec{\chi}}_j^\mathrm{copr}(t_i)$ and
$\vec{\Omega}^\mathrm{copr}(t_i)$ by
rotating the $x$ and $y$ components of the corresponding coorbital
quantities by an angle of $-\varphi(t_i)$.
We evolve the coprecessing vectors instead of the coorbital vectors
because the former evolve on the longer precession timescale, allowing
us to take large timesteps.
Finally, after computing
\begin{equation}
    \left.\frac{d}{dt}\hat{q}(t)\right\rvert_{t_i} =
        \frac{1}{2}\hat{q}(t_i)\vec{\Omega}^\mathrm{copr}(t_i)\,,
\end{equation}
we obtain the time derivatives of $\varphi$, $\hat{q}$,
and $\vec{\chi}_j^\mathrm{copr}$ at $t=t_i$.  

These time derivatives are then used to integrate $\varphi$, $\hat{q}$,
and $\vec{\chi}_j^\mathrm{copr}$ using an ODE solver.
We desire an ODE integration method that uses
few evaluations of the time derivatives to keep the computational
cost of evaluating the model low.
We use a fourth-order Adams-Bashforth method~\cite{Butcher2003,BA66330495}
detailed in Appendix~\ref{app:ab4},
which determines the solutions at the next node based on the time derivatives
at the current and three previous nodes.
This allows us to reuse fit evaluations from the previous nodes, and requires
only one additional evaluation of the fits per node compared to four evaluations
for a fourth-order Runge-Kutta scheme.
The Adams-Bashforth integration is initialized by performing the first
integration
steps with fourth-order Runge-Kutta.
This is why we include the three
additional time nodes
$t_\frac{1}{2}$, $t_\frac{3}{2}$ and $t_\frac{5}{2}$; they
enable evaluating the midpoint increments of the initial Runge-Kutta scheme.
Once we have evaluated the solutions at the time nodes $t_i$, we use
cubic spline interpolation to determine the solutions
at all times.

Now that we have $\varphi$, $\hat{q}$,
and $\vec{\chi}_j^\mathrm{copr}$ for all $t$, we
then evaluate each coorbital waveform data piece.
This is done by first evaluating the fits at the empirical nodes $T_i$
using the mass ratio $q$ and the coorbital spins at the empirical nodes
$\vec{\chi}_j^\mathrm{coorb}(T_i)$, and then evaluating the empirical
interpolant to obtain the waveform data piece at all times.
Finally, we transform the coorbital frame waveform modes back to the
coprecessing frame using $\varphi(t)$ and then to the
inertial frame using $\hat{q}(t)$.
The NRSur7dq2 surrogate data and Python evaluation code can be found
at~\cite{SpECSurrogates}.

To reduce the computational cost of transforming the coprecessing waveform
modes to the inertial frame using $\hat{q}(t)$, which takes $\sim~1\,\mathrm{s}$
using all $\ell \leq 4$ modes sampled with $\delta t=0.1M$, we reduce the
number of time samples of the coorbital waveform data pieces by using
non-uniform time steps.
We choose $2000$ time samples that are roughly uniformly spaced in the
orbital phase, using the same method used to choose the dynamics time
nodes described in Appendix~\ref{app:timesamples}.
This is sufficiently many time samples to yield negligible errors
when interpolating back to the dense uniformly-spaced time array
using cubic splines on the real and imaginary parts of the waveform modes.

Integrating the ODE system takes $\sim~3\,\mathrm{ms}$, where the numerical
computations are performed by a Python extension written in C.
Interpolating the results of the ODE integration to the $2000$ time samples
described above takes $\sim~2\,\mathrm{ms}$ using cubic splines.
Evaluating the coorbital waveform surrogate takes $\sim~4\,\mathrm{ms}$,
and transforming the modes to the inertial frame takes $\sim~16\,\mathrm{ms}$,
for a total of $\sim~25\,\mathrm{ms}$.
Variations in the evaluation time can increase this
up to $\sim~30\,\mathrm{ms}$.
Restricting to only $\ell=2$ modes can reduce this time
to $\sim~10\,\mathrm{ms}$.
  If we wish to sample the surrogate waveform at the same time nodes
  as the original
  numerical relativity simulations, which is a uniformly-spaced time array
with $\delta t = 0.1M$, the modes are
interpolated to these points using
cubic splines. This requires
$\sim~6\,\mathrm{ms}$ per mode, for a total time of
$\sim~150\,\mathrm{ms}$ when all $\ell \leq 4$ modes are
interpolated in this way.
We note, however, that the original NR simulations are
oversampled for typical GW
data analysis purposes.
For example, a sampling rate of $4096\,\mathrm{Hz}$ for a $M=60M_\odot$
binary has $\delta t \approx 0.83M$, leading to an evaluation time of
$\sim~50\,\mathrm{ms}$.
All timings were done on Intel Xeon E5-2680v3 cores running
  at 2.5GHz.

%%%%%%%%%%%%%%%%%%%%%%%%%%%%%%%%%%%%%%%%%%%%%%%%%%%%%%%%%%%%%%%%%%%%%%%%%%%%%%%
\section{Surrogate Errors}
%%%%%%%%%%%%%%%%%%%%%%%%%%%%%%%%%%%%%%%%%%%%%%%%%%%%%%%%%%%%%%%%%%%%%%%%%%%%%%%

We use two error measures to quantify the accuracy of the surrogate model.
Given two sets of waveform modes $h_1$ and $h_2$, we first compute
\begin{equation}
\label{eq:mathcalE}
\mathcal{E}[h_1, h_2] = \frac{1}{2} \frac{
    \sum_{\ell, m} \int_{t_0}^{t_f} |h_1^{\ell, m}(t) - h_2^{\ell, m}(t)|^2 dt}{
    \sum_{\ell, m} \int_{t_0}^{t_f} |h_1^{\ell, m}(t)|^2 dt}\,,
\end{equation}
which is introduced in Eq.~(21) of~\cite{Blackman:2017dfb}.
Since we have aligned all the NR waveforms at $t=t_0$ and the surrogate
model reproduces this alignment, we do not perform any time or phase shifts
when computing $\mathcal{E}$.

For these comparisons, we use modes $\ell \le 5$;
  if a mode is not included in a particular waveform model,
  we assume this mode is zero for that model. Since the NRSur7dq2 model does not contain
  $\ell=5$ modes, this ensures that the errors discussed below include
the effect of neglecting $\ell=5$ and higher modes.

Histograms of $\mathcal{E}$ for all $886$ NR waveforms
are given in Fig.~\ref{fig:component_errors}.
For all curves in the figure, $h_1$ is the
highest available resolution NR waveform.
For the thick solid black curve, $h_2$ is the same
NR waveform as $h_1$, except computed at a lower numerical resolution,
so this curve represents an estimate of the numerical
truncation error in the NR waveforms used to build the surrogate model.
For the solid blue curve, $h_2$ is the  NRSur7dq2 surrogate waveform evaluated with the same mass ratio and initial spins of $h_1$.
Note that since the surrogate was trained using all NR waveforms, this
is an \emph{in-sample} error.

The remaining curves in Fig.~\ref{fig:component_errors}
indicate the in-sample error contribution from each of the
three main waveform data pieces in the surrogate waveform:
the orbital phase $\varphi$
(dash-dotted green curve),
the quaternions $\hat{q}$ representing the precession (dashed orange curve),
and the waveform modes
in a coorbital frame $h_\mathrm{coorb}$ (thin solid red curve).
For these curves, $h_2$ is computed by
using the surrogate evaluation for 
one waveform data piece and the NR
evaluation of the other pieces.
The orbital phase errors give rise to the largest surrogate errors,
indicating that efforts to improve the surrogate model should be
focused on improving the orbital phasing.

\begin{figure}
    \includegraphics[width=\linewidth]{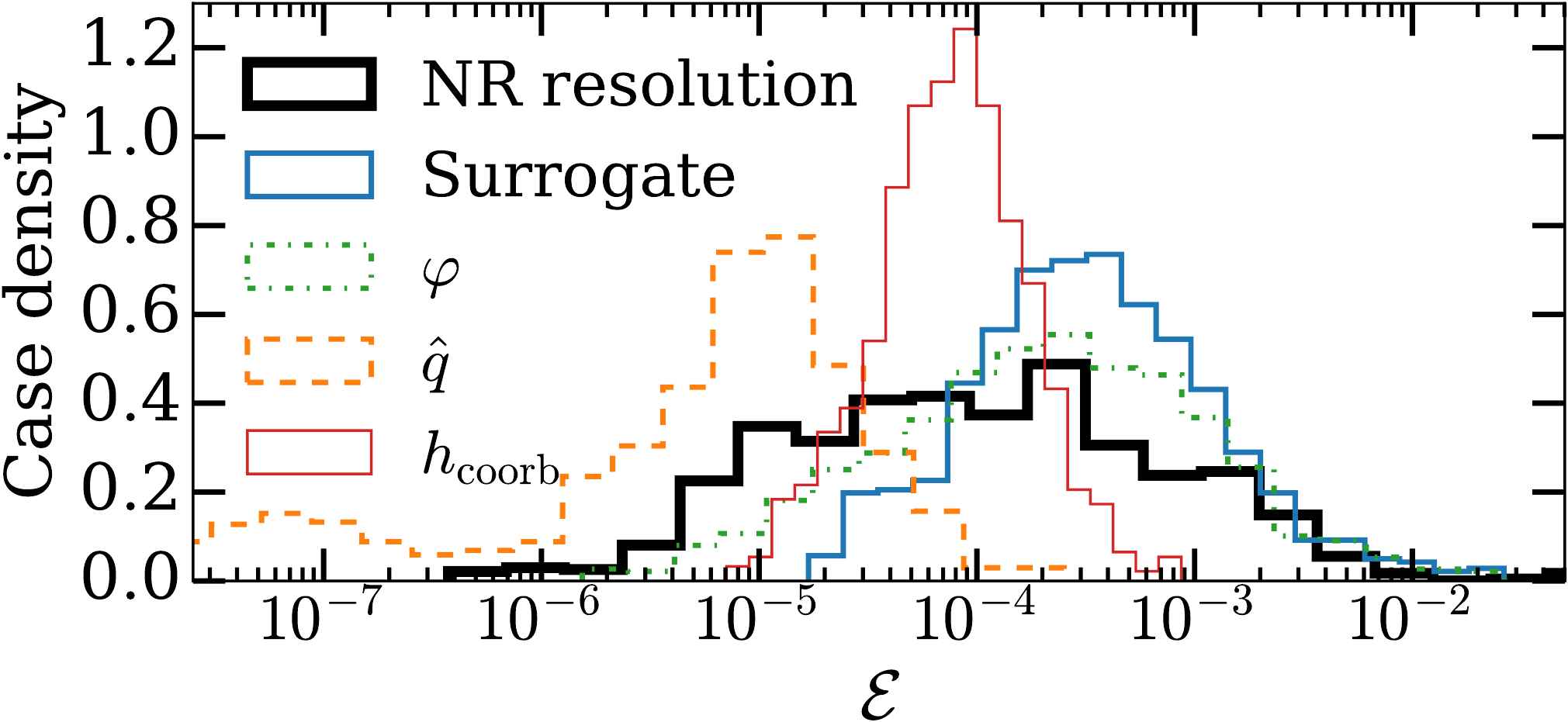}
    \caption{
        Error histograms for $\mathcal{E}$ defined in
        Eq.~(\ref{eq:mathcalE}), normalized such that the area under
        each curve is $1$ when integrated over $\mathrm{log}_{10}(\mathcal{E})$.
        The largest surrogate errors are comparable to the largest NR resolution
        errors, which compare high and medium resolution NR simulations to
        estimate the error in the NR waveforms.
        The error in the orbital
        phase $\varphi$ is the dominant error contribution to
        the surrogate.
    }
    \label{fig:component_errors}
\end{figure}

We then compute mismatches~\cite{Babak:2016tgq}
\begin{equation}
1 - \frac{\langle h_1, h_2\rangle}{\sqrt{\langle h_1, h_1\rangle
                                         \langle h_2, h_2\rangle}}\,,
\end{equation}
where $\langle \cdot, \cdot \rangle$ is a noise-weighted inner product
computed in the frequency domain, as in Sec.~VI.B
of Ref.~\cite{Blackman:2017dfb}.
We use a flat
power spectral density to avoid a dependence on the total mass of the system.
The mismatches are minimized over timeshifts, polarization angle shifts,
and shifts in the azimuthal angle of the direction of GW
propagation, where the system's orbital angular momentum is initially
aligned with the $\hat{z}$ axis. We randomly sample $30$ directions
of gravitational wave propagation on the sphere, and use a pair
of detectors with idealized orientations such that one detector
measures $h_+$ and the other detector measures $h_\times$.
Histograms of the mismatches are given in Fig.~\ref{fig:mismatches} and are
comparable to the top panel of Fig.~$17$ in Ref.~\cite{Blackman:2017dfb}.
To estimate the \emph{out-of-sample} errors of the surrogate model,
we perform a $20$-fold cross-validation test.
This is done by first randomly dividing the $886$ NR waveforms into $20$
sets of $44$ or $45$ waveforms.
For each set, we build a \emph{trial surrogate} using the waveforms
from the other $19$ sets.
The trial surrogate is then evaluated at the parameters corresponding to the
waveforms in the chosen \emph{validation} set, and the results are compared to the
NR waveform.
These cross-validation mismatches are given by the dashed purple curve.
They are quite similar to the in-sample errors given by the solid blue curve,
indicating that we are not overfitting the data.
We
also compute mismatches for
a fully precessing effective one-body model
(SEOBNRv3~\cite{Pan:2013rra}), and for a phenomenological waveform model
that includes some,
but not all, effects of precession (IMRPhenomPv2~\cite{Hannam:2013oca}).
These models have mismatches more than an order of magnitude larger
than our NRSur7dq2 surrogate model.
Both IMRPhenomPv2 and SEOBNRv3 depend on a parameter
  $f_\mathrm{ref}$, which is a reference
  frequency at which the spin directions are specified.
  For SEOBNRv3, which is a time-domain model, we
  choose $f_\mathrm{ref}$ so that the waveform begins at $t=t_0$.
  For IMRPhenomPv2, which is a frequency-domain model, we minimize
  the mismatches over $f_\mathrm{ref}$, using an initial guess of twice
  the orbital frequency of the NR waveform at $t=t_0$.
While all of the mismatches can be decreased by minimizing over additional
parameters such as BH masses and spins,
this would result in biased parameters when measuring
the source parameters of a detected GW signal.

\begin{figure}
    \includegraphics[width=\linewidth]{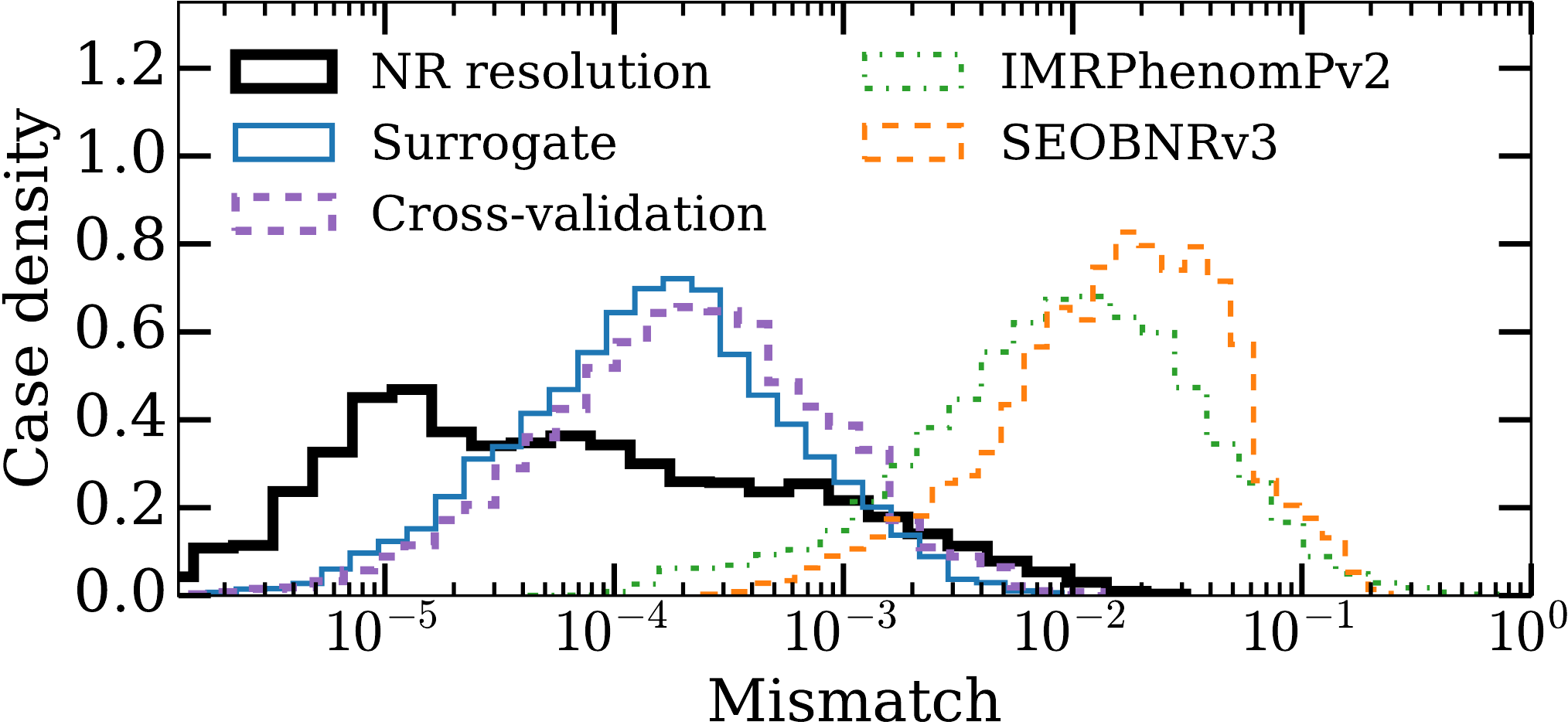}
    \caption{
        Mismatch histograms computed in the frequency domain with a flat
        power spectral density.
        The NR resolution mismatches compare waveforms from high and
        medium resolution NR simulations. This can be an overestimate
        of the error in the high resolution NR waveform, leading to
        some NR resolution mismatches being larger than the surrogate
        mismatches.
        We note that the IMRPhenomPv2 model does not contain all spin
        components.
    }
    \label{fig:mismatches}
\end{figure}

We then compute mismatches using the advanced LIGO design sensitivity
noise curve~\cite{Shoemaker2009, aLIGO2} using various total masses $M$.
For each mass $M$, we obtain histograms as in Fig.~\ref{fig:mismatches},
and we show the median and $95$th percentile mismatches from these histograms
in Fig.~\ref{fig:mismatch_vs_mass}.
We note that for $M \lesssim 114M_\odot$ some or all waveforms begin above
$10\,\mathrm{Hz}$
and do not cover the full design sensitivity frequency band.
We find that the $95$th percentile mismatches of our surrogate model
are similar to the corresponding NR mismatches, except for total masses above
$160M_\odot$ where the NR mismatches are slightly smaller.
The NRSur7dq2 surrogate yields mismatches at least an order of magnitude
smaller than the other waveform models for all total masses investigated.

\begin{figure}
    \includegraphics[width=\linewidth]{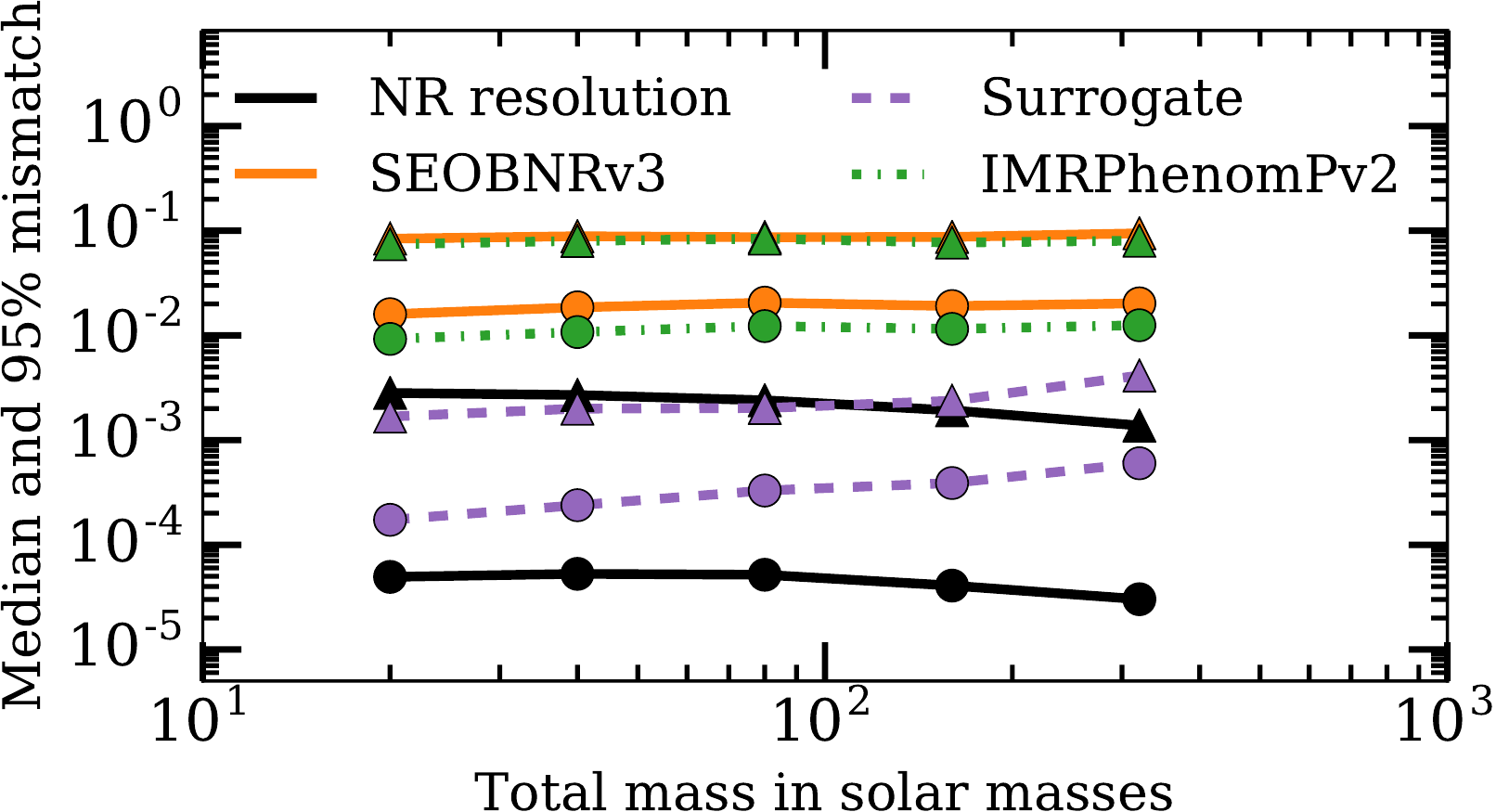}
    \caption{
        Median (circles) and $95$th percentile (triangles) mismatches
        of all $866$ cases computed with the advanced LIGO design
        sensitivity curve.
        The surrogate mismatches are computed using trial surrogates,
        as in the cross-validation curve of Fig.~\ref{fig:mismatches}.
    }
    \label{fig:mismatch_vs_mass}
\end{figure}

Figure~\ref{fig:waveforms} shows
the real part of $h^{2, 2}(t)$ for the cases leading to the largest
mismatches in Fig.~\ref{fig:mismatches}.
The top panel shows the case leading to the largest surrogate cross-validation
mismatch, and the bottom panel shows the case leading to the largest
SEOBNRv3 mismatch.
The surrogate waveforms shown are evaluated using the appropriate trial surrogate, so that they were not trained on the NR waveforms they are compared with.
All waveforms are aligned to have their peak amplitude at $t=0$
and are rotated to have their orbital angular momentum aligned with
the $z$ axis at $t=t_0=-4500M$.
In the top panel, we see that both the SEOBNRv3 and surrogate waveforms
have a similar phasing error around $t=-50M$.
The phasing error of the surrogate does not grow significantly
larger through merger and ringdown, so most of this error can be removed
with a time and phase shift.
For the SEOBNRv3 waveforms in both the upper and lower panels,
the phasing error changes
significantly during the merger; therefore
this error does not decrease significantly even after
performing a time and phase shift.  In the top panel of
Figure~\ref{fig:waveforms}, the IMRPhenomPv2 waveform does as
well as the surrogate; in the bottom panel, the IMRPhenomPv2 waveform
has large errors in both phase and amplitude.

\begin{figure}
    \includegraphics[width=\linewidth]{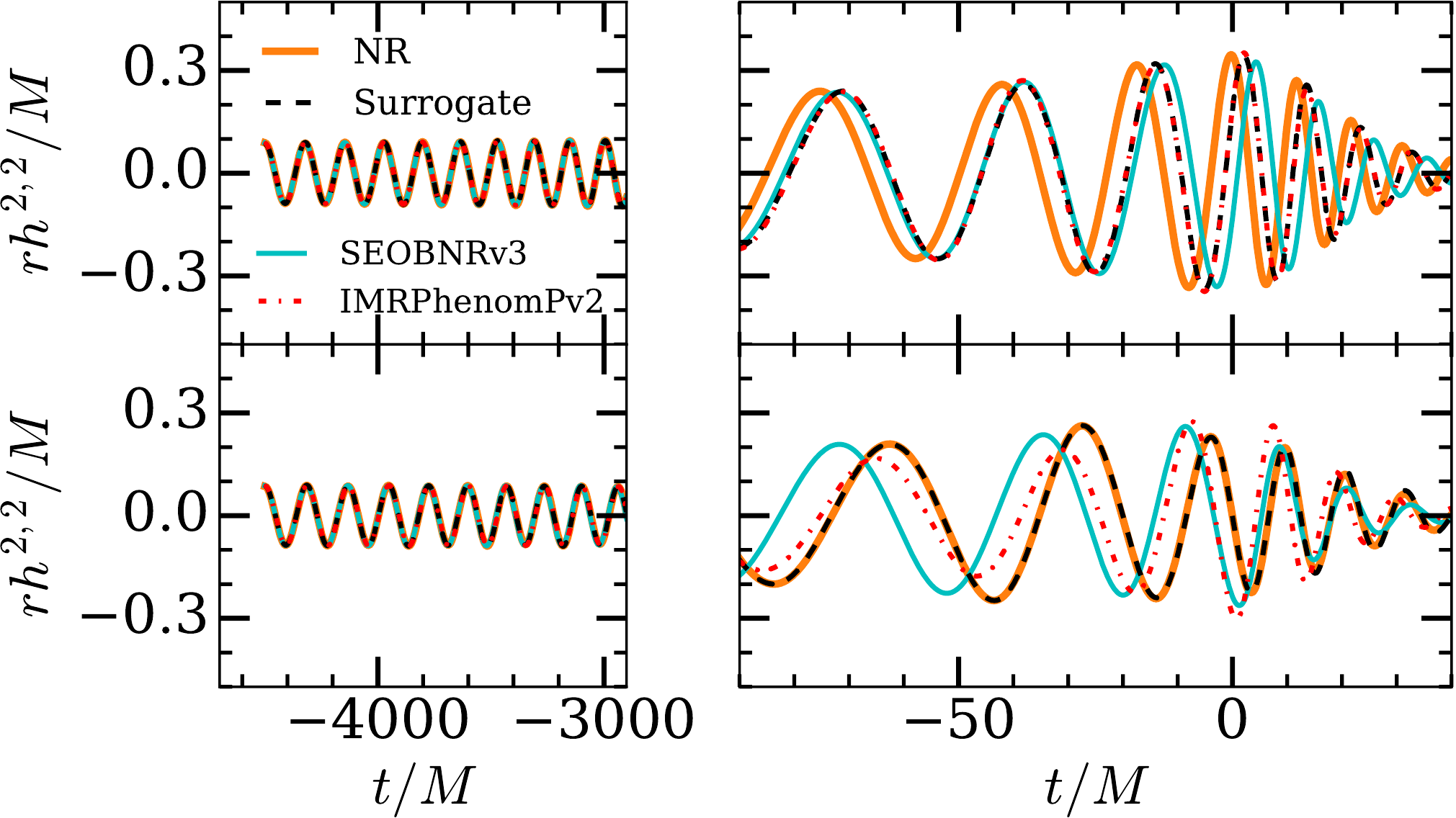}
    \caption{
        The real part of time domain waveforms
        for the case leading to the largest
        surrogate mismatch (top) and the largest SEOBNRv3 mismatch (bottom).
        The surrogate waveforms are evaluated using trial surrogates
        which were not trained with the NR waveform shown.
        The top panel uses SXS:BBH:0922 with $q\approx2$,
        $\vec{\chi}_1(t_0) \approx 0.8\hat{z}$, and
        $\vec{\chi}_2(t_0) \approx -0.8\hat{y}$.
        The lower panel uses SXS:BBH:0900 with $q\approx2$,
        $\vec{\chi}_1(t_0) \approx (0.29, -0.74, 0.02)$ and
        $\vec{\chi}_2(t_0) \approx (0.43, -0.34, 0.58)$.
        %with a time and phase shift.
        %The SEOBNRv3 phasing error changes significantly near merger,
        %leading to large errors even after a time and phase shift.
    }
    \label{fig:waveforms}
\end{figure}

\section{Discussion and Conclusions}

Within its range of validity, our NRSur7dq2 surrogate model is
nearly as accurate as performing new NR simulations.
The surrogate model takes only
$\sim~50\,\mathrm{ms}$ to evaluate on a single CPU core,
making it sufficiently fast for current GW data analysis
applications such as parameter estimation.
This evaluation time can be compared to $\mathcal{O}(\mathrm{weeks})$
on dozens of CPU cores to perform a new NR simulation, decreasing the
cost in CPU-hours by $\mathcal{O}(10^8)$.
The NRSur7dq2 surrogate model data along with Python evaluation code
is publicly available for download
at~\cite{SpECSurrogates}.

Our surrogate model is limited to mass ratios $q \leq 2$ and spin
magnitudes $|\vec{\chi_{1,2}}| \leq 0.8$.
While in principle the parametric fits can be extrapolated
to more extreme mass ratios and spin magnitudes, we do
not expect extrapolation to
  yield accurate waveforms.
However, these
  limits can be extended in future versions of our surrogate model
by performing NR simulations with larger mass ratios and spins.

Additionally, the waveforms produced by NRSur7dq2 are limited in
duration to $4500M$ before the peak amplitude.
This covers frequencies $f \geq 20\,\mathrm{Hz}$
for all systems with $M \gtrsim 57 M_\odot$.
For systems with lower total masses, or for systems
with $M \lesssim 114 M_\odot$
when including frequencies down to $10\,\mathrm{Hz}$,
longer waveforms are needed.
In future work, we
plan to
overcome this limitation by hybridizing with either PN or
SEOBNRv3~\cite{Bustillo:2015ova, Boyle:2011dy,
        OhmeEtAl:2011, MacDonald:2011ne},
either by hybridizing the NR waveforms before building the surrogate
or by hybridizing the surrogate waveforms.
Longer NR waveforms would then be needed to test the accuracy of the hybridization
step.

%%%%%%%%%%%%%%%%%%%%%%%%%%%%%%%%%%%%%%%%%%%%%%%%%%%%%%%%%%%%%%%%%%%%%%%%%%%%%%%
% Acknowledgments
%%%%%%%%%%%%%%%%%%%%%%%%%%%%%%%%%%%%%%%%%%%%%%%%%%%%%%%%%%%%%%%%%%%%%%%%%%%%%%%
\begin{acknowledgments}
We thank Matt Giesler for helping to carry out the new SpEC simulations
used in this work.
We thank Saul Teukolsky, Patricia Schmidt, Rory Smith, and Vijay Varma for helpful
discussions.
This work was supported in part by the Sherman Fairchild Foundation
and by NSF grants CAREER PHY-1151197, PHY-1404569,
AST-1333129, and PHY-1606654.
Computations were performed on NSF/NCSA Blue Waters under allocation
PRAC ACI-1440083;
on the NSF XSEDE network under allocation TG-PHY100033; and on the Zwicky cluster at
Caltech, which is supported by the Sherman Fairchild Foundation and by NSF
award PHY-0960291.
This paper has been assigned YITP report number YITP-17-44.
\end{acknowledgments}

%%%%%%%%%%%%%%%%%%%%%%%%%%%%%%%%%%%%%%%%%%%%%%%%%%%%%%%%%%%%%%%%%%%%%%%%%%%%%%%
\appendix

\section{Sparse grid parameters}
\label{app:parameters}
We take the polar and azimuthal spin angles of the inertial
frame spins $\vec{\chi}_i$
to be $\theta_i$ and $\phi_i$ respectively, for $i \in \{1, 2\}$.
We can then parametrize our $7$-dimensional parameter space by
\begin{itemize}
    \item $q \in [1, 2]$,
    \item $|\vec{\chi}_i| \in [0, 0.8]$,
    \item $\theta_i \in [0, \pi]$,
    \item $\phi_i \in [0, 2\pi]$.
\end{itemize}
The range of each of these variables is some closed interval $[a, b]$.
For a variable $x$ with range $[a, b]$, we define a grid of $N$ uniformly-spaced
points
\begin{equation}
g_x^N = \left\{a + \frac{n}{N-1}(b-a) : n=0,\ldots,N-1\right\}\,,
\end{equation}
where $N \geq 2$.
We then define a sequence of grids
\begin{equation}
G_x \equiv G_x^0,\,G_x^1,\,\ldots\,,
\end{equation}
where
\begin{equation}
G_x^n = g_x^{f_x(n)}
\end{equation}
for some monotonically increasing function $f_x(n)$.
We call $G_x^n$ the \emph{level $n$ grid for $x$}.
We take
\begin{align}
f_q(n) &= f_{|\vec{\chi_i}|}(n) = 1 + 2^n,\\
f_{\theta_i}(n) &= 1 + 2^{n+1},\\
f_{\phi_i}(n) &= 1 + 3 \cdot 2^n\,.
\end{align}
These choices ensure that $G_x^n \subset G_x^{n+1}$,
and that the level $0$ grids already give a description of the
parameter space that does not leave out any phenomenology;
the level $0$ grids for $\theta_i$ contain the midpoint
$\pi/2$ leading to precession, and the level $0$ grids for
$\phi_i$ contain $3$ unique points (since $\phi_i=0$ and
$\phi_i=2\pi$ lead to the same physical spin)
in order to get at least some resolution of features that
behave like $\mathrm{sin}(\phi_i + \phi+0)$.

We have already seen that $\phi_i=0$ and $\phi_i=2\pi$
correspond to the same physical spin, but we will have
many other scenarios where two combinations of variables
lead to the same physical configuration.
For example, if $|\vec{\chi}_1| =0$, all combinations
of $\theta_1$ and $\phi_1$ lead to the same physical
configuration.
We will ignore these degenerate combinations for now,
and remove them later on.

Dense grids in parameter space could be constructed as
\begin{align*}
G_\mathrm{dense}^n = G_q^n \times
    & G_{|\vec{\chi}_1|}^n \times G_{\theta_1}^n \times G_{\phi_1}^n \times \\
    & G_{|\vec{\chi}_2|}^n \times G_{\theta_2}^n \times G_{\phi_2}^n\,,
\end{align*}
where $\times$ denotes the Cartesian product.
While the $1$-dimensional grids grow in size as $\mathcal{O}(2^n)$,
these dense grids grow in size as $\mathcal{O}(2^{7n})$ or as the seventh
power of the size of the $1$-dimensional grids.
This is known as the \emph{curse of dimensionality};
the amount of data needed often grows exponentially with the
dimensionality.
Sparse grids~\cite{smolyak1963quadrature, bungartz2004sparse} overcome the curse of
dimensionality by using a sparse product
such that the grids grow in size as 
%$\mathcal{O}(2^n)$.
$\mathcal{O}\left(2^n \left(\log 2^n \right)^{6} \right)$.
If $G_{x}$ and $G_{y}$ are two sequences of grids,
we define the \emph{sparse product} of $G_{x}$ with $G_{y}$ to be
$G_{x, y} = G_{x} \bullet G_{y}$,
where
\begin{equation}
G_{x, y}^n = \bigcup\limits_{k=0}^n G_x^k \times G_y^{n - k}\,.
\end{equation}

We now define the sparse grids for our parameter space from
the sequence of grids
\begin{equation}
G = G_q \bullet G_{|\vec{\chi}_1|} \bullet G_{\theta_1} \bullet G_{\phi_1}
    \bullet G_{|\vec{\chi}_2|} \bullet G_{\theta_2} \bullet G_{\phi_2}
\end{equation}
such that
\begin{align*}
G^n = \bigcup\limits_{\sum_{i=1}^7 k_i = n} G_q^{k_1} &\times
    G_{|\vec{\chi}_1|}^{k_2} \times G_{\theta_1}^{k_3} \times G_{\phi_1}^{k_4}\\
        & \times
    G_{|\vec{\chi}_2|}^{k_5} \times G_{\theta_2}^{k_6} \times G_{\phi_2}^{k_7}\,.
\end{align*}

%We performed $361$ new NR simulations based on parameters in $G^1$.
%We removed physically identical configurations.
%We also removed configurations with $\vec{\chi}_2 \propto \hat{z}$,
%which are within the parameter space of the NRSur4d2s surrogate model,
%which was already covered by the $276$ NRSur4d2s NR simulations.

Starting with the parameters in $G^1$ we
removed physically identical configurations. We also removed 
configurations with $\vec{\chi}_2 \propto \hat{z}$,
which are within the parameter space of the NRSur4d2s surrogate model,
which was already covered by the $276$ NRSur4d2s NR simulations.
We performed $361$ new NR simulations based on the remaining set of parameter 
values. 

%%%%%%%%%%%%%%%%%%%%%%%%%%%%%%%%%%%%%%%%%%%%%%%%%%%%%%%%%%%%%%%%%%%%%%%%%%%%%%%
\section{Time sampling}
\label{app:timesamples}

We wish to choose time nodes $t_0 < t_1 < \ldots < t_f$
that are roughly uniformly spaced in the
orbital phase $\varphi(t)$ for all cases.
Given some number $N$, we choose time nodes yielding roughly $N$ nodes per
orbit.
Since different NR waveforms have different orbital frequencies, they will
have a different number of time nodes per orbit.
Our scheme for choosing the time nodes given $N$ is based on the leading order
PN expression for the orbital angular frequency $\omega(t)$
during the inspiral,
smoothly transitioning
to a maximum value of $\omega = 2\pi/(20M)$ 
during the ringdown.
We do this by computing a bounded time
\begin{equation}
\tilde{t}(t) = -1.7 + \frac{1}{2}\left((t+5) - \sqrt{(t+5)^2 + 25}\right) \,,
\end{equation}
and then choosing
\begin{equation}
\omega_\mathrm{ref}(t) = \omega_\mathrm{0PN}(\tilde{t}(t)) = \left(
        \frac{64}{5}(-\tilde{t}(t))\right)^{-\frac{3}{8}}\,.
\end{equation}
We then use spacings between nodes $t_{j+1} - t_j = \omega_\mathrm{ref}(t_j)$.

%%%%%%%%%%%%%%%%%%%%%%%%%%%%%%%%%%%%%%%%%%%%%%%%%%%%%%%%%%%%%%%%%%%%%%%%%%%%%%%
\section{Fourth-order Adams-Bashforth method}
\label{app:ab4}

We integrate the ODE system on a non-uniformly spaced grid of time nodes
$t_0 < t_1 < \ldots < t_f$ using a fourth-order Adams-Bashforth
scheme~\cite{Butcher2003,BA66330495}.
We denote the solution $\vec{y}(t)$, and at each time node $t_i$, we can
evaluate fits to determine
\begin{equation}
\frac{d\vec{y}}{dt} = \vec{f}(t; \vec{y})\,.
\end{equation}
We first integrate up to $t_3$ using a Runge-Kutta fourth-order scheme.

Once we have integrated up to $t_i$ for $i >= 3$,
we have previously evaluated
\begin{equation}
\vec{k}_j = \vec{f}(t_j; \vec{y}(t_j))
\end{equation}
for $0 \leq j < i$, and we now evaluate $\vec{k}_i$.
We approximate $\vec{g}(t) = \vec{f}(t; \vec{y}(t))$
by a cubic function
\begin{equation}
\vec{g}(t) \approx \vec{g}_3(t) = \vec{A} + \vec{B}(t - t_i) + \vec{C}(t - t_i)^2 + \vec{D}(t - t_i)^3\,.
\end{equation}
The coefficients are chosen such that $\vec{g}_3(t_j) = \vec{g}(t_j) = \vec{k}_j$
for $i-3 \leq j \leq i$, giving $\vec{A} = \vec{k}_i$, and
\begin{equation*}
\begin{bmatrix}
\vec{B} \\
\vec{C} \\
\vec{D} \\
\end{bmatrix} = \begin{bmatrix}
\frac{\delta_{-1, 0}\delta_{-2, 0}}{\Delta_1} &
    \frac{\delta_{-1, 0}\delta_{-3, 0}}{\Delta_2} &
    \frac{\delta_{-2, 0}\delta_{-3, 0}}{\Delta_3} \\
\frac{\delta_{-2, 0} + \delta_{-1, 0}}{\Delta_1} &
    \frac{\delta_{-1, 0} + \delta_{-3, 0}}{\Delta_2} &
    \frac{\delta_{-2, 0} + \delta_{-3, 0}}{\Delta_3} \\
\frac{1}{\Delta_1} &
     \frac{1}{\Delta_2} &
     \frac{1}{\Delta_3}
\end{bmatrix} \begin{bmatrix}
\vec{k}_i - \vec{k}_{i-3} \\
\vec{k}_i - \vec{k}_{i-2} \\
\vec{k}_i - \vec{k}_{i-1}
\end{bmatrix}\,.
\end{equation*}
Here, $\delta_{n, m} = t_{i+m} - t_{i+n}$ and
\begin{align}
\Delta_1 &= \delta_{-3, -2}\delta_{-3, -1}\delta_{-3, 0}\,, \\
\Delta_2 &= \delta_{-3, -2}\delta_{-2, -1}\delta_{-2, 0}\,, \\
\Delta_3 &= \delta_{-2, -1}\delta_{-3, -1}\delta_{-1, 0}\,.
\end{align}
Finally, we approximate
\begin{align*}
\vec{y}(t_{i+1}) &= \vec{y}(t_i) + \int_{t_i}^{t_{i+1}} g(t) dt \\
                 &\approx \vec{y}(t_i) + \int_{t_i}^{t_{i+1}} g_3(t_j) dt \\
                 &=  \vec{y}(t_i) + \delta_{0, 1}\vec{A} +
                      \frac{1}{2}\delta_{0, 1}^2\vec{B} +
                      \frac{1}{3}\delta_{0, 1}^3\vec{C} +
                      \frac{1}{4}\delta_{0, 1}^4\vec{D}\,.
\end{align*}
%%%%%%%%%%%%%%%%%%%%%%%%%%%%%%%%%%%%%%%%%%%%%%%%%%%%%%%%%%%%%%%%%%%%%%%%%%%%%%%
\section*{References}
%%%%%%%%%%%%%%%%%%%%%%%%%%%%%%%%%%%%%%%%%%%%%%%%%%%%%%%%%%%%%%%%%%%%%%%%%%%%%%%
\bibliography{References/References}

%merlin.mbs apsrev4-1.bst 2010-07-25 4.21a (PWD, AO, DPC) hacked
%Control: key (0)
%Control: author (8) initials jnrlst
%Control: editor formatted (1) identically to author
%Control: production of article title (-1) disabled
%Control: page (0) single
%Control: year (1) truncated
%Control: production of eprint (0) enabled
\begin{thebibliography}{56}%
\makeatletter
\providecommand \@ifxundefined [1]{%
 \@ifx{#1\undefined}
}%
\providecommand \@ifnum [1]{%
 \ifnum #1\expandafter \@firstoftwo
 \else \expandafter \@secondoftwo
 \fi
}%
\providecommand \@ifx [1]{%
 \ifx #1\expandafter \@firstoftwo
 \else \expandafter \@secondoftwo
 \fi
}%
\providecommand \natexlab [1]{#1}%
\providecommand \enquote  [1]{``#1''}%
\providecommand \bibnamefont  [1]{#1}%
\providecommand \bibfnamefont [1]{#1}%
\providecommand \citenamefont [1]{#1}%
\providecommand \href@noop [0]{\@secondoftwo}%
\providecommand \href [0]{\begingroup \@sanitize@url \@href}%
\providecommand \@href[1]{\@@startlink{#1}\@@href}%
\providecommand \@@href[1]{\endgroup#1\@@endlink}%
\providecommand \@sanitize@url [0]{\catcode `\\12\catcode `\$12\catcode
  `\&12\catcode `\#12\catcode `\^12\catcode `\_12\catcode `\%12\relax}%
\providecommand \@@startlink[1]{}%
\providecommand \@@endlink[0]{}%
\providecommand \url  [0]{\begingroup\@sanitize@url \@url }%
\providecommand \@url [1]{\endgroup\@href {#1}{\urlprefix }}%
\providecommand \urlprefix  [0]{URL }%
\providecommand \Eprint [0]{\href }%
\providecommand \doibase [0]{http://dx.doi.org/}%
\providecommand \selectlanguage [0]{\@gobble}%
\providecommand \bibinfo  [0]{\@secondoftwo}%
\providecommand \bibfield  [0]{\@secondoftwo}%
\providecommand \translation [1]{[#1]}%
\providecommand \BibitemOpen [0]{}%
\providecommand \bibitemStop [0]{}%
\providecommand \bibitemNoStop [0]{.\EOS\space}%
\providecommand \EOS [0]{\spacefactor3000\relax}%
\providecommand \BibitemShut  [1]{\csname bibitem#1\endcsname}%
\let\auto@bib@innerbib\@empty
%</preamble>
\bibitem [{\citenamefont {Abbott}\ \emph
  {et~al.}(2016{\natexlab{a}})\citenamefont {Abbott} \emph
  {et~al.}}]{LIGOVirgo2016a}%
  \BibitemOpen
  \bibfield  {author} {\bibinfo {author} {\bibfnamefont {B.~P.}\ \bibnamefont
  {Abbott}} \emph {et~al.} (\bibinfo {collaboration} {LIGO Scientific
  Collaboration, Virgo Collaboration}),\ }\href {\doibase
  10.1103/PhysRevLett.116.061102} {\bibfield  {journal} {\bibinfo  {journal}
  {Phys.\ Rev.\ Lett.}\ }\textbf {\bibinfo {volume} {116}},\ \bibinfo {pages}
  {061102} (\bibinfo {year} {2016}{\natexlab{a}})},\ \Eprint
  {http://arxiv.org/abs/1602.03837} {arXiv:1602.03837 [gr-qc]} \BibitemShut
  {NoStop}%
\bibitem [{\citenamefont {Abbott}\ \emph
  {et~al.}(2016{\natexlab{b}})\citenamefont {Abbott} \emph
  {et~al.}}]{Abbott:2016nmj}%
  \BibitemOpen
  \bibfield  {author} {\bibinfo {author} {\bibfnamefont {B.~P.}\ \bibnamefont
  {Abbott}} \emph {et~al.} (\bibinfo {collaboration} {LIGO Scientific
  Collaboration, Virgo Collaboration}),\ }\href {\doibase
  10.1103/PhysRevLett.116.241103} {\bibfield  {journal} {\bibinfo  {journal}
  {Phys. Rev. Lett.}\ }\textbf {\bibinfo {volume} {116}},\ \bibinfo {pages}
  {241103} (\bibinfo {year} {2016}{\natexlab{b}})},\ \Eprint
  {http://arxiv.org/abs/1606.04855} {arXiv:1606.04855 [gr-qc]} \BibitemShut
  {NoStop}%
%%CITATION = ARXIV:1606.04855;%%
\bibitem [{\citenamefont {Abbott}\ \emph
  {et~al.}(2016{\natexlab{c}})\citenamefont {Abbott} \emph
  {et~al.}}]{TheLIGOScientific:2016wfe}%
  \BibitemOpen
  \bibfield  {author} {\bibinfo {author} {\bibfnamefont {B.~P.}\ \bibnamefont
  {Abbott}} \emph {et~al.} (\bibinfo {collaboration} {LIGO Scientific
  Collaboration, Virgo Collaboration}),\ }\href {\doibase
  10.1103/PhysRevLett.116.241102} {\bibfield  {journal} {\bibinfo  {journal}
  {Phys.~Rev.~Lett.}\ }\textbf {\bibinfo {volume} {116}},\ \bibinfo {pages}
  {241102} (\bibinfo {year} {2016}{\natexlab{c}})},\ \Eprint
  {http://arxiv.org/abs/1602.03840} {arXiv:1602.03840 [gr-qc]} \BibitemShut
  {NoStop}%
%%CITATION = ARXIV:1602.03840;%%
\bibitem [{\citenamefont {Abbott}\ \emph
  {et~al.}(2016{\natexlab{d}})\citenamefont {Abbott} \emph
  {et~al.}}]{Abbott:2016apu}%
  \BibitemOpen
  \bibfield  {author} {\bibinfo {author} {\bibfnamefont {B.~P.}\ \bibnamefont
  {Abbott}} \emph {et~al.} (\bibinfo {collaboration} {LIGO Scientific
  Collaboration, Virgo Collaboration}),\ }\href@noop {} {\bibfield  {journal}
  {\bibinfo  {journal} {Phys.~Rev.~D}\ }\textbf {\bibinfo {volume} {94}},\
  \bibinfo {pages} {064035} (\bibinfo {year} {2016}{\natexlab{d}})},\ \Eprint
  {http://arxiv.org/abs/1606.01262} {arXiv:1606.01262 [gr-qc]} \BibitemShut
  {NoStop}%
%%CITATION = ARXIV:1606.01262;%%
\bibitem [{\citenamefont {Abbott}\ \emph
  {et~al.}(2016{\natexlab{e}})\citenamefont {Abbott} \emph
  {et~al.}}]{TheLIGOScientific:2016src}%
  \BibitemOpen
  \bibfield  {author} {\bibinfo {author} {\bibfnamefont {B.~P.}\ \bibnamefont
  {Abbott}} \emph {et~al.} (\bibinfo {collaboration} {LIGO Scientific
  Collaboration, Virgo Collaboration}),\ }\href@noop {} {\bibfield  {journal}
  {\bibinfo  {journal} {Phys.~Rev.~Lett.}\ }\textbf {\bibinfo {volume} {116}},\
  \bibinfo {pages} {221101} (\bibinfo {year} {2016}{\natexlab{e}})},\ \Eprint
  {http://arxiv.org/abs/1602.03841} {arXiv:1602.03841 [gr-qc]} \BibitemShut
  {NoStop}%
%%CITATION = ARXIV:1602.03841;%%
\bibitem [{\citenamefont {Pretorius}(2005)}]{Pretorius2005a}%
  \BibitemOpen
  \bibfield  {author} {\bibinfo {author} {\bibfnamefont {F.}~\bibnamefont
  {Pretorius}},\ }\href {\doibase 10.1103/PhysRevLett.95.121101} {\bibfield
  {journal} {\bibinfo  {journal} {Phys.\ Rev.\ Lett.}\ }\textbf {\bibinfo
  {volume} {95}},\ \bibinfo {pages} {121101} (\bibinfo {year} {2005})},\
  \Eprint {http://arxiv.org/abs/gr-qc/0507014} {arXiv:gr-qc/0507014 [gr-qc]}
  \BibitemShut {NoStop}%
%%CITATION = GR-QC/0507014;%%
\bibitem [{\citenamefont {Zlochower}\ \emph {et~al.}(2005)\citenamefont
  {Zlochower}, \citenamefont {Baker}, \citenamefont {Campanelli},\ and\
  \citenamefont {Lousto}}]{Zlochower:2005bj}%
  \BibitemOpen
  \bibfield  {author} {\bibinfo {author} {\bibfnamefont {Y.}~\bibnamefont
  {Zlochower}}, \bibinfo {author} {\bibfnamefont {J.}~\bibnamefont {Baker}},
  \bibinfo {author} {\bibfnamefont {M.}~\bibnamefont {Campanelli}}, \ and\
  \bibinfo {author} {\bibfnamefont {C.}~\bibnamefont {Lousto}},\ }\href
  {\doibase 10.1103/PhysRevD.72.024021} {\bibfield  {journal} {\bibinfo
  {journal} {Phys.\ Rev.\ D}\ }\textbf {\bibinfo {volume} {72}},\ \bibinfo
  {pages} {024021} (\bibinfo {year} {2005})},\ \Eprint
  {http://arxiv.org/abs/gr-qc/0505055} {arXiv:gr-qc/0505055 [gr-qc]}
  \BibitemShut {NoStop}%
%%CITATION = GR-QC/0505055;%%
\bibitem [{SpE({\natexlab{a}})}]{SpECwebsite}%
  \BibitemOpen
  \href@noop {} {}\bibinfo {howpublished}
  {\url{http://www.black-holes.org/SpEC.html}} ({\natexlab{a}})\BibitemShut
  {NoStop}%
\bibitem [{ein()}]{einsteintoolkit}%
  \BibitemOpen
  \href@noop {} {}\bibinfo {note} {Einstein Toolkit home page:
  \url{http://einsteintoolkit.org}}\BibitemShut {NoStop}%
\bibitem [{\citenamefont {Husa}\ \emph {et~al.}(2008)\citenamefont {Husa},
  \citenamefont {Gonz{\'a}lez}, \citenamefont {Hannam}, \citenamefont
  {Br{\"u}gmann},\ and\ \citenamefont {Sperhake}}]{Husa2007}%
  \BibitemOpen
  \bibfield  {author} {\bibinfo {author} {\bibfnamefont {S.}~\bibnamefont
  {Husa}}, \bibinfo {author} {\bibfnamefont {J.~A.}\ \bibnamefont
  {Gonz{\'a}lez}}, \bibinfo {author} {\bibfnamefont {M.}~\bibnamefont
  {Hannam}}, \bibinfo {author} {\bibfnamefont {B.}~\bibnamefont
  {Br{\"u}gmann}}, \ and\ \bibinfo {author} {\bibfnamefont {U.}~\bibnamefont
  {Sperhake}},\ }\href@noop {} {\bibfield  {journal} {\bibinfo  {journal}
  {Class.\ Quantum Grav.}\ }\textbf {\bibinfo {volume} {25}},\ \bibinfo {pages}
  {105006} (\bibinfo {year} {2008})}\BibitemShut {NoStop}%
\bibitem [{\citenamefont {Br{\"u}gmann}\ \emph {et~al.}(2008)\citenamefont
  {Br{\"u}gmann}, \citenamefont {Gonz\'{a}lez}, \citenamefont {Hannam},
  \citenamefont {Husa}, \citenamefont {Sperhake},\ and\ \citenamefont
  {Tichy}}]{Bruegmann2006}%
  \BibitemOpen
  \bibfield  {author} {\bibinfo {author} {\bibfnamefont {B.}~\bibnamefont
  {Br{\"u}gmann}}, \bibinfo {author} {\bibfnamefont {J.~A.}\ \bibnamefont
  {Gonz\'{a}lez}}, \bibinfo {author} {\bibfnamefont {M.}~\bibnamefont
  {Hannam}}, \bibinfo {author} {\bibfnamefont {S.}~\bibnamefont {Husa}},
  \bibinfo {author} {\bibfnamefont {U.}~\bibnamefont {Sperhake}}, \ and\
  \bibinfo {author} {\bibfnamefont {W.}~\bibnamefont {Tichy}},\ }\href
  {\doibase 10.1103/PhysRevD.77.024027} {\bibfield  {journal} {\bibinfo
  {journal} {Phys.\ Rev.\ D}\ }\textbf {\bibinfo {volume} {77}},\ \bibinfo
  {eid} {024027} (\bibinfo {year} {2008})},\ \Eprint
  {http://arxiv.org/abs/gr-qc/0610128} {gr-qc/0610128} \BibitemShut {NoStop}%
\bibitem [{\citenamefont {Herrmann}\ \emph {et~al.}(2007)\citenamefont
  {Herrmann}, \citenamefont {Hinder}, \citenamefont {Shoemaker},\ and\
  \citenamefont {Laguna}}]{Herrmann2007b}%
  \BibitemOpen
  \bibfield  {author} {\bibinfo {author} {\bibfnamefont {F.}~\bibnamefont
  {Herrmann}}, \bibinfo {author} {\bibfnamefont {I.}~\bibnamefont {Hinder}},
  \bibinfo {author} {\bibfnamefont {D.}~\bibnamefont {Shoemaker}}, \ and\
  \bibinfo {author} {\bibfnamefont {P.}~\bibnamefont {Laguna}},\ }\href@noop {}
  {\bibfield  {journal} {\bibinfo  {journal} {Class.\ Quantum Grav.}\ }\textbf
  {\bibinfo {volume} {24}},\ \bibinfo {pages} {S33} (\bibinfo {year} {2007})},\
  \Eprint {http://arxiv.org/abs/gr-qc/0601026} {gr-qc/0601026} \BibitemShut
  {NoStop}%
\bibitem [{\citenamefont {Damour}\ \emph {et~al.}(2008)\citenamefont {Damour},
  \citenamefont {Jaranowski},\ and\ \citenamefont {Schaefer}}]{Damour:024009}%
  \BibitemOpen
  \bibfield  {author} {\bibinfo {author} {\bibfnamefont {T.}~\bibnamefont
  {Damour}}, \bibinfo {author} {\bibfnamefont {P.}~\bibnamefont {Jaranowski}},
  \ and\ \bibinfo {author} {\bibfnamefont {G.}~\bibnamefont {Schaefer}},\
  }\href {\doibase 10.1103/PhysRevD.78.024009} {\bibfield  {journal} {\bibinfo
  {journal} {Phys.\ Rev.\ D}\ }\textbf {\bibinfo {volume} {78}},\ \bibinfo
  {pages} {024009} (\bibinfo {year} {2008})},\ \Eprint
  {http://arxiv.org/abs/0803.0915} {arXiv:0803.0915 [gr-qc]} \BibitemShut
  {NoStop}%
%%CITATION = ARXIV:0803.0915;%%
\bibitem [{\citenamefont {Damour}\ and\ \citenamefont
  {Nagar}(2009)}]{Damour2009a}%
  \BibitemOpen
  \bibfield  {author} {\bibinfo {author} {\bibfnamefont {T.}~\bibnamefont
  {Damour}}\ and\ \bibinfo {author} {\bibfnamefont {A.}~\bibnamefont {Nagar}},\
  }\href {\doibase 10.1103/PhysRevD.79.081503} {\bibfield  {journal} {\bibinfo
  {journal} {Phys.\ Rev.\ D}\ }\textbf {\bibinfo {volume} {79}},\ \bibinfo
  {pages} {081503} (\bibinfo {year} {2009})},\ \Eprint
  {http://arxiv.org/abs/0902.0136} {arXiv:0902.0136 [gr-qc]} \BibitemShut
  {NoStop}%
%%CITATION = ARXIV:0902.0136;%%
\bibitem [{\citenamefont {Taracchini}\ \emph {et~al.}(2014)\citenamefont
  {Taracchini}, \citenamefont {Buonanno}, \citenamefont {Pan}, \citenamefont
  {Hinderer}, \citenamefont {Boyle}, \citenamefont {Hemberger}, \citenamefont
  {Kidder}, \citenamefont {Lovelace}, \citenamefont {Mroue}, \citenamefont
  {Pfeiffer}, \citenamefont {Scheel}, \citenamefont {Szil{\'a}gyi},
  \citenamefont {Taylor},\ and\ \citenamefont
  {Zenginoglu}}]{Taracchini:2013rva}%
  \BibitemOpen
  \bibfield  {author} {\bibinfo {author} {\bibfnamefont {A.}~\bibnamefont
  {Taracchini}}, \bibinfo {author} {\bibfnamefont {A.}~\bibnamefont
  {Buonanno}}, \bibinfo {author} {\bibfnamefont {Y.}~\bibnamefont {Pan}},
  \bibinfo {author} {\bibfnamefont {T.}~\bibnamefont {Hinderer}}, \bibinfo
  {author} {\bibfnamefont {M.}~\bibnamefont {Boyle}}, \bibinfo {author}
  {\bibfnamefont {D.~A.}\ \bibnamefont {Hemberger}}, \bibinfo {author}
  {\bibfnamefont {L.~E.}\ \bibnamefont {Kidder}}, \bibinfo {author}
  {\bibfnamefont {G.}~\bibnamefont {Lovelace}}, \bibinfo {author}
  {\bibfnamefont {A.~H.}\ \bibnamefont {Mroue}}, \bibinfo {author}
  {\bibfnamefont {H.~P.}\ \bibnamefont {Pfeiffer}}, \bibinfo {author}
  {\bibfnamefont {M.~A.}\ \bibnamefont {Scheel}}, \bibinfo {author}
  {\bibfnamefont {B.}~\bibnamefont {Szil{\'a}gyi}}, \bibinfo {author}
  {\bibfnamefont {N.~W.}\ \bibnamefont {Taylor}}, \ and\ \bibinfo {author}
  {\bibfnamefont {A.}~\bibnamefont {Zenginoglu}},\ }\href@noop {} {\bibfield
  {journal} {\bibinfo  {journal} {Phys.\ Rev.\ D}\ }\textbf {\bibinfo {volume}
  {89 (R)}},\ \bibinfo {pages} {061502} (\bibinfo {year} {2014})},\ \Eprint
  {http://arxiv.org/abs/1311.2544} {arXiv:1311.2544 [gr-qc]} \BibitemShut
  {NoStop}%
\bibitem [{\citenamefont {{P{\"u}rrer}}(2016)}]{Purrer:2015tud}%
  \BibitemOpen
  \bibfield  {author} {\bibinfo {author} {\bibfnamefont {M.}~\bibnamefont
  {{P{\"u}rrer}}},\ }\href {\doibase 10.1103/PhysRevD.93.064041} {\bibfield
  {journal} {\bibinfo  {journal} {Phys.\ Rev.\ D}\ }\textbf {\bibinfo {volume}
  {93}},\ \bibinfo {eid} {064041} (\bibinfo {year} {2016})},\ \Eprint
  {http://arxiv.org/abs/1512.02248} {arXiv:1512.02248 [gr-qc]} \BibitemShut
  {NoStop}%
\bibitem [{\citenamefont {{Pan}}\ \emph {et~al.}(2013)\citenamefont {{Pan}},
  \citenamefont {{Buonanno}}, \citenamefont {{Taracchini}}, \citenamefont
  {{Kidder}}, \citenamefont {{Mrou{\'e}}}, \citenamefont {{Pfeiffer}},
  \citenamefont {{Scheel}},\ and\ \citenamefont
  {{Szil{\'a}gyi}}}]{Pan:2013rra}%
  \BibitemOpen
  \bibfield  {author} {\bibinfo {author} {\bibfnamefont {Y.}~\bibnamefont
  {{Pan}}}, \bibinfo {author} {\bibfnamefont {A.}~\bibnamefont {{Buonanno}}},
  \bibinfo {author} {\bibfnamefont {A.}~\bibnamefont {{Taracchini}}}, \bibinfo
  {author} {\bibfnamefont {L.~E.}\ \bibnamefont {{Kidder}}}, \bibinfo {author}
  {\bibfnamefont {A.~H.}\ \bibnamefont {{Mrou{\'e}}}}, \bibinfo {author}
  {\bibfnamefont {H.~P.}\ \bibnamefont {{Pfeiffer}}}, \bibinfo {author}
  {\bibfnamefont {M.~A.}\ \bibnamefont {{Scheel}}}, \ and\ \bibinfo {author}
  {\bibfnamefont {B.}~\bibnamefont {{Szil{\'a}gyi}}},\ }\href@noop {}
  {\bibfield  {journal} {\bibinfo  {journal} {Phys.\ Rev.\ D}\ }\textbf
  {\bibinfo {volume} {89}},\ \bibinfo {pages} {084006} (\bibinfo {year}
  {2013})},\ \Eprint {http://arxiv.org/abs/1307.6232} {arXiv:1307.6232 [gr-qc]}
  \BibitemShut {NoStop}%
%%CITATION = ARXIV:1307.6232;%%
\bibitem [{\citenamefont {Boh\'e}\ \emph {et~al.}(2017)\citenamefont {Boh\'e},
  \citenamefont {Shao}, \citenamefont {Taracchini}, \citenamefont {Buonanno},
  \citenamefont {Babak}, \citenamefont {Harry}, \citenamefont {Hinder},
  \citenamefont {Ossokine}, \citenamefont {P\"urrer}, \citenamefont {Raymond},
  \citenamefont {Chu}, \citenamefont {Fong}, \citenamefont {Kumar},
  \citenamefont {Pfeiffer}, \citenamefont {Boyle}, \citenamefont {Hemberger},
  \citenamefont {Kidder}, \citenamefont {Lovelace}, \citenamefont {Scheel},\
  and\ \citenamefont {Szil\'agyi}}]{Bohe:2016gbl}%
  \BibitemOpen
  \bibfield  {author} {\bibinfo {author} {\bibfnamefont {A.}~\bibnamefont
  {Boh\'e}}, \bibinfo {author} {\bibfnamefont {L.}~\bibnamefont {Shao}},
  \bibinfo {author} {\bibfnamefont {A.}~\bibnamefont {Taracchini}}, \bibinfo
  {author} {\bibfnamefont {A.}~\bibnamefont {Buonanno}}, \bibinfo {author}
  {\bibfnamefont {S.}~\bibnamefont {Babak}}, \bibinfo {author} {\bibfnamefont
  {I.~W.}\ \bibnamefont {Harry}}, \bibinfo {author} {\bibfnamefont
  {I.}~\bibnamefont {Hinder}}, \bibinfo {author} {\bibfnamefont
  {S.}~\bibnamefont {Ossokine}}, \bibinfo {author} {\bibfnamefont
  {M.}~\bibnamefont {P\"urrer}}, \bibinfo {author} {\bibfnamefont
  {V.}~\bibnamefont {Raymond}}, \bibinfo {author} {\bibfnamefont
  {T.}~\bibnamefont {Chu}}, \bibinfo {author} {\bibfnamefont {H.}~\bibnamefont
  {Fong}}, \bibinfo {author} {\bibfnamefont {P.}~\bibnamefont {Kumar}},
  \bibinfo {author} {\bibfnamefont {H.~P.}\ \bibnamefont {Pfeiffer}}, \bibinfo
  {author} {\bibfnamefont {M.}~\bibnamefont {Boyle}}, \bibinfo {author}
  {\bibfnamefont {D.~A.}\ \bibnamefont {Hemberger}}, \bibinfo {author}
  {\bibfnamefont {L.~E.}\ \bibnamefont {Kidder}}, \bibinfo {author}
  {\bibfnamefont {G.}~\bibnamefont {Lovelace}}, \bibinfo {author}
  {\bibfnamefont {M.~A.}\ \bibnamefont {Scheel}}, \ and\ \bibinfo {author}
  {\bibfnamefont {B.}~\bibnamefont {Szil\'agyi}},\ }\href {\doibase
  10.1103/PhysRevD.95.044028} {\bibfield  {journal} {\bibinfo  {journal} {Phys.
  Rev. D}\ }\textbf {\bibinfo {volume} {95}},\ \bibinfo {pages} {044028}
  (\bibinfo {year} {2017})},\ \Eprint {http://arxiv.org/abs/1611.03703}
  {arXiv:1611.03703 [gr-qc]} \BibitemShut {NoStop}%
%%CITATION = ARXIV:1611.03703;%%
\bibitem [{\citenamefont {Hannam}\ \emph {et~al.}(2014)\citenamefont {Hannam},
  \citenamefont {Schmidt}, \citenamefont {Boh{\' e}}, \citenamefont {Haegel},
  \citenamefont {Husa} \emph {et~al.}}]{Hannam:2013oca}%
  \BibitemOpen
  \bibfield  {author} {\bibinfo {author} {\bibfnamefont {M.}~\bibnamefont
  {Hannam}}, \bibinfo {author} {\bibfnamefont {P.}~\bibnamefont {Schmidt}},
  \bibinfo {author} {\bibfnamefont {A.}~\bibnamefont {Boh{\' e}}}, \bibinfo
  {author} {\bibfnamefont {L.}~\bibnamefont {Haegel}}, \bibinfo {author}
  {\bibfnamefont {S.}~\bibnamefont {Husa}},  \emph {et~al.},\ }\href {\doibase
  10.1103/PhysRevLett.113.151101} {\bibfield  {journal} {\bibinfo  {journal}
  {Phys.\ Rev.\ Lett.}\ }\textbf {\bibinfo {volume} {113}},\ \bibinfo {pages}
  {151101} (\bibinfo {year} {2014})},\ \Eprint {http://arxiv.org/abs/1308.3271}
  {arXiv:1308.3271 [gr-qc]} \BibitemShut {NoStop}%
%%CITATION = ARXIV:1308.3271;%%
\bibitem [{\citenamefont {Khan}\ \emph {et~al.}(2016)\citenamefont {Khan},
  \citenamefont {Husa}, \citenamefont {Hannam}, \citenamefont {Ohme},
  \citenamefont {Pürrer}, \citenamefont {Jiménez~Forteza},\ and\
  \citenamefont {Bohé}}]{Khan:2015jqa}%
  \BibitemOpen
  \bibfield  {author} {\bibinfo {author} {\bibfnamefont {S.}~\bibnamefont
  {Khan}}, \bibinfo {author} {\bibfnamefont {S.}~\bibnamefont {Husa}}, \bibinfo
  {author} {\bibfnamefont {M.}~\bibnamefont {Hannam}}, \bibinfo {author}
  {\bibfnamefont {F.}~\bibnamefont {Ohme}}, \bibinfo {author} {\bibfnamefont
  {M.}~\bibnamefont {Pürrer}}, \bibinfo {author} {\bibfnamefont
  {X.}~\bibnamefont {Jiménez~Forteza}}, \ and\ \bibinfo {author}
  {\bibfnamefont {A.}~\bibnamefont {Bohé}},\ }\href {\doibase
  10.1103/PhysRevD.93.044007} {\bibfield  {journal} {\bibinfo  {journal} {Phys.
  Rev.}\ }\textbf {\bibinfo {volume} {D93}},\ \bibinfo {pages} {044007}
  (\bibinfo {year} {2016})},\ \Eprint {http://arxiv.org/abs/1508.07253}
  {arXiv:1508.07253 [gr-qc]} \BibitemShut {NoStop}%
%%CITATION = ARXIV:1508.07253;%%
\bibitem [{\citenamefont {Husa}\ \emph {et~al.}(2016)\citenamefont {Husa},
  \citenamefont {Khan}, \citenamefont {Hannam}, \citenamefont {Pürrer},
  \citenamefont {Ohme}, \citenamefont {Jim\'{e}nez~Forteza},\ and\
  \citenamefont {Boh\'{e}}}]{Husa:2015iqa}%
  \BibitemOpen
  \bibfield  {author} {\bibinfo {author} {\bibfnamefont {S.}~\bibnamefont
  {Husa}}, \bibinfo {author} {\bibfnamefont {S.}~\bibnamefont {Khan}}, \bibinfo
  {author} {\bibfnamefont {M.}~\bibnamefont {Hannam}}, \bibinfo {author}
  {\bibfnamefont {M.}~\bibnamefont {Pürrer}}, \bibinfo {author} {\bibfnamefont
  {F.}~\bibnamefont {Ohme}}, \bibinfo {author} {\bibfnamefont {X.}~\bibnamefont
  {Jim\'{e}nez~Forteza}}, \ and\ \bibinfo {author} {\bibfnamefont
  {A.}~\bibnamefont {Boh\'{e}}},\ }\href {\doibase 10.1103/PhysRevD.93.044006}
  {\bibfield  {journal} {\bibinfo  {journal} {Phys. Rev.}\ }\textbf {\bibinfo
  {volume} {D93}},\ \bibinfo {pages} {044006} (\bibinfo {year} {2016})},\
  \Eprint {http://arxiv.org/abs/1508.07250} {arXiv:1508.07250 [gr-qc]}
  \BibitemShut {NoStop}%
%%CITATION = ARXIV:1508.07250;%%
\bibitem [{\citenamefont {Lindblom}\ \emph {et~al.}(2008)\citenamefont
  {Lindblom}, \citenamefont {Owen},\ and\ \citenamefont
  {Brown}}]{Lindblom2008}%
  \BibitemOpen
  \bibfield  {author} {\bibinfo {author} {\bibfnamefont {L.}~\bibnamefont
  {Lindblom}}, \bibinfo {author} {\bibfnamefont {B.~J.}\ \bibnamefont {Owen}},
  \ and\ \bibinfo {author} {\bibfnamefont {D.~A.}\ \bibnamefont {Brown}},\
  }\href {\doibase 10.1103/PhysRevD.78.124020} {\bibfield  {journal} {\bibinfo
  {journal} {Phys.\ Rev.\ D}\ }\textbf {\bibinfo {volume} {78}},\ \bibinfo
  {pages} {124020} (\bibinfo {year} {2008})},\ \Eprint
  {http://arxiv.org/abs/0809.3844} {arXiv:0809.3844 [gr-qc]} \BibitemShut
  {NoStop}%
%%CITATION = 0809.3844;%%
\bibitem [{\citenamefont {Abbott}\ \emph
  {et~al.}(2016{\natexlab{f}})\citenamefont {Abbott} \emph
  {et~al.}}]{Abbott:2016wiq}%
  \BibitemOpen
  \bibfield  {author} {\bibinfo {author} {\bibfnamefont {B.~P.}\ \bibnamefont
  {Abbott}} \emph {et~al.} (\bibinfo {collaboration} {Virgo, LIGO
  Scientific}),\ }\href@noop {} {\  (\bibinfo {year} {2016}{\natexlab{f}})},\
  \Eprint {http://arxiv.org/abs/1611.07531} {arXiv:1611.07531 [gr-qc]}
  \BibitemShut {NoStop}%
%%CITATION = ARXIV:1611.07531;%%
\bibitem [{\citenamefont {Blackman}\ \emph {et~al.}(2017)\citenamefont
  {Blackman}, \citenamefont {Field}, \citenamefont {Scheel}, \citenamefont
  {Galley}, \citenamefont {Hemberger}, \citenamefont {Schmidt},\ and\
  \citenamefont {Smith}}]{Blackman:2017dfb}%
  \BibitemOpen
  \bibfield  {author} {\bibinfo {author} {\bibfnamefont {J.}~\bibnamefont
  {Blackman}}, \bibinfo {author} {\bibfnamefont {S.~E.}\ \bibnamefont {Field}},
  \bibinfo {author} {\bibfnamefont {M.~A.}\ \bibnamefont {Scheel}}, \bibinfo
  {author} {\bibfnamefont {C.~R.}\ \bibnamefont {Galley}}, \bibinfo {author}
  {\bibfnamefont {D.~A.}\ \bibnamefont {Hemberger}}, \bibinfo {author}
  {\bibfnamefont {P.}~\bibnamefont {Schmidt}}, \ and\ \bibinfo {author}
  {\bibfnamefont {R.}~\bibnamefont {Smith}},\ }\href@noop {} {\  (\bibinfo
  {year} {2017})},\ \Eprint {http://arxiv.org/abs/1701.00550} {arXiv:1701.00550
  [gr-qc]} \BibitemShut {NoStop}%
%%CITATION = ARXIV:1701.00550;%%
\bibitem [{\citenamefont {{Blackman}}\ \emph {et~al.}(2015)\citenamefont
  {{Blackman}}, \citenamefont {{Field}}, \citenamefont {{Galley}},
  \citenamefont {{Szil{\'a}gyi}}, \citenamefont {{Scheel}}, \citenamefont
  {{Tiglio}},\ and\ \citenamefont {{Hemberger}}}]{Blackman:2015pia}%
  \BibitemOpen
  \bibfield  {author} {\bibinfo {author} {\bibfnamefont {J.}~\bibnamefont
  {{Blackman}}}, \bibinfo {author} {\bibfnamefont {S.~E.}\ \bibnamefont
  {{Field}}}, \bibinfo {author} {\bibfnamefont {C.~R.}\ \bibnamefont
  {{Galley}}}, \bibinfo {author} {\bibfnamefont {B.}~\bibnamefont
  {{Szil{\'a}gyi}}}, \bibinfo {author} {\bibfnamefont {M.~A.}\ \bibnamefont
  {{Scheel}}}, \bibinfo {author} {\bibfnamefont {M.}~\bibnamefont {{Tiglio}}},
  \ and\ \bibinfo {author} {\bibfnamefont {D.~A.}\ \bibnamefont
  {{Hemberger}}},\ }\href {\doibase 10.1103/PhysRevLett.115.121102} {\bibfield
  {journal} {\bibinfo  {journal} {Phys.\ Rev.\ Lett.}\ }\textbf {\bibinfo
  {volume} {115}},\ \bibinfo {eid} {121102} (\bibinfo {year} {2015})},\ \Eprint
  {http://arxiv.org/abs/1502.07758} {arXiv:1502.07758 [gr-qc]} \BibitemShut
  {NoStop}%
\bibitem [{\citenamefont {{Field}}\ \emph {et~al.}(2014)\citenamefont
  {{Field}}, \citenamefont {{Galley}}, \citenamefont {{Hesthaven}},
  \citenamefont {{Kaye}},\ and\ \citenamefont {{Tiglio}}}]{Field:2013cfa}%
  \BibitemOpen
  \bibfield  {author} {\bibinfo {author} {\bibfnamefont {S.~E.}\ \bibnamefont
  {{Field}}}, \bibinfo {author} {\bibfnamefont {C.~R.}\ \bibnamefont
  {{Galley}}}, \bibinfo {author} {\bibfnamefont {J.~S.}\ \bibnamefont
  {{Hesthaven}}}, \bibinfo {author} {\bibfnamefont {J.}~\bibnamefont {{Kaye}}},
  \ and\ \bibinfo {author} {\bibfnamefont {M.}~\bibnamefont {{Tiglio}}},\
  }\href {\doibase 10.1103/PhysRevX.4.031006} {\bibfield  {journal} {\bibinfo
  {journal} {Phys.\ Rev.\ X}\ }\textbf {\bibinfo {volume} {4}},\ \bibinfo {eid}
  {031006} (\bibinfo {year} {2014})},\ \Eprint {http://arxiv.org/abs/1308.3565}
  {arXiv:1308.3565 [gr-qc]} \BibitemShut {NoStop}%
\bibitem [{\citenamefont {P{\"u}rrer}(2014)}]{Purrer:2014}%
  \BibitemOpen
  \bibfield  {author} {\bibinfo {author} {\bibfnamefont {M.}~\bibnamefont
  {P{\"u}rrer}},\ }\href {http://stacks.iop.org/0264-9381/31/i=19/a=195010}
  {\bibfield  {journal} {\bibinfo  {journal} {Class.\ Quantum Grav.}\ }\textbf
  {\bibinfo {volume} {31}},\ \bibinfo {pages} {195010} (\bibinfo {year}
  {2014})},\ \Eprint {http://arxiv.org/abs/1402.4146} {arXiv:1402.4146 [gr-qc]}
  \BibitemShut {NoStop}%
\bibitem [{\citenamefont {Pfeiffer}\ \emph {et~al.}(2003)\citenamefont
  {Pfeiffer}, \citenamefont {Kidder}, \citenamefont {Scheel},\ and\
  \citenamefont {Teukolsky}}]{Pfeiffer2003}%
  \BibitemOpen
  \bibfield  {author} {\bibinfo {author} {\bibfnamefont {H.~P.}\ \bibnamefont
  {Pfeiffer}}, \bibinfo {author} {\bibfnamefont {L.~E.}\ \bibnamefont
  {Kidder}}, \bibinfo {author} {\bibfnamefont {M.~A.}\ \bibnamefont {Scheel}},
  \ and\ \bibinfo {author} {\bibfnamefont {S.~A.}\ \bibnamefont {Teukolsky}},\
  }\href {\doibase 10.1016/S0010-4655(02)00847-0} {\bibfield  {journal}
  {\bibinfo  {journal} {Comput.\ Phys.\ Commun.}\ }\textbf {\bibinfo {volume}
  {152}},\ \bibinfo {pages} {253} (\bibinfo {year} {2003})},\ \Eprint
  {http://arxiv.org/abs/gr-qc/0202096} {gr-qc/0202096} \BibitemShut {NoStop}%
\bibitem [{\citenamefont {Lovelace}\ \emph {et~al.}(2008)\citenamefont
  {Lovelace}, \citenamefont {Owen}, \citenamefont {Pfeiffer},\ and\
  \citenamefont {Chu}}]{Lovelace2008}%
  \BibitemOpen
  \bibfield  {author} {\bibinfo {author} {\bibfnamefont {G.}~\bibnamefont
  {Lovelace}}, \bibinfo {author} {\bibfnamefont {R.}~\bibnamefont {Owen}},
  \bibinfo {author} {\bibfnamefont {H.~P.}\ \bibnamefont {Pfeiffer}}, \ and\
  \bibinfo {author} {\bibfnamefont {T.}~\bibnamefont {Chu}},\ }\href {\doibase
  10.1103/PhysRevD.78.084017} {\bibfield  {journal} {\bibinfo  {journal}
  {Phys.\ Rev.\ D}\ }\textbf {\bibinfo {volume} {78}},\ \bibinfo {pages}
  {084017} (\bibinfo {year} {2008})}\BibitemShut {NoStop}%
\bibitem [{\citenamefont {Lindblom}\ \emph {et~al.}(2006)\citenamefont
  {Lindblom}, \citenamefont {Scheel}, \citenamefont {Kidder}, \citenamefont
  {Owen},\ and\ \citenamefont {Rinne}}]{Lindblom2006}%
  \BibitemOpen
  \bibfield  {author} {\bibinfo {author} {\bibfnamefont {L.}~\bibnamefont
  {Lindblom}}, \bibinfo {author} {\bibfnamefont {M.~A.}\ \bibnamefont
  {Scheel}}, \bibinfo {author} {\bibfnamefont {L.~E.}\ \bibnamefont {Kidder}},
  \bibinfo {author} {\bibfnamefont {R.}~\bibnamefont {Owen}}, \ and\ \bibinfo
  {author} {\bibfnamefont {O.}~\bibnamefont {Rinne}},\ }\href {\doibase
  10.1088/0264-9381/23/16/S09} {\bibfield  {journal} {\bibinfo  {journal}
  {Class.\ Quantum Grav.}\ }\textbf {\bibinfo {volume} {23}},\ \bibinfo {pages}
  {S447} (\bibinfo {year} {2006})},\ \Eprint
  {http://arxiv.org/abs/gr-qc/0512093} {gr-qc/0512093} \BibitemShut {NoStop}%
\bibitem [{\citenamefont {{Szil{\'a}gyi}}\ \emph {et~al.}(2009)\citenamefont
  {{Szil{\'a}gyi}}, \citenamefont {{Lindblom}},\ and\ \citenamefont
  {{Scheel}}}]{Szilagyi:2009qz}%
  \BibitemOpen
  \bibfield  {author} {\bibinfo {author} {\bibfnamefont {B.}~\bibnamefont
  {{Szil{\'a}gyi}}}, \bibinfo {author} {\bibfnamefont {L.}~\bibnamefont
  {{Lindblom}}}, \ and\ \bibinfo {author} {\bibfnamefont {M.~A.}\ \bibnamefont
  {{Scheel}}},\ }\href {\doibase 10.1103/PhysRevD.80.124010} {\bibfield
  {journal} {\bibinfo  {journal} {Phys.\ Rev.\ D}\ }\textbf {\bibinfo {volume}
  {80}},\ \bibinfo {eid} {124010} (\bibinfo {year} {2009})},\ \Eprint
  {http://arxiv.org/abs/0909.3557} {arXiv:0909.3557 [gr-qc]} \BibitemShut
  {NoStop}%
\bibitem [{\citenamefont {{M. A. Scheel, M. Boyle, T. Chu, L. E. Kidder, K. D.
  Matthews and H. P. Pfeiffer}}(2009)}]{Scheel2009}%
  \BibitemOpen
  \bibfield  {author} {\bibinfo {author} {\bibnamefont {{M. A. Scheel, M.
  Boyle, T. Chu, L. E. Kidder, K. D. Matthews and H. P. Pfeiffer}}},\
  }\href@noop {} {\bibfield  {journal} {\bibinfo  {journal} {Phys.\ Rev.\ D}\
  }\textbf {\bibinfo {volume} {79}},\ \bibinfo {pages} {024003} (\bibinfo
  {year} {2009})},\ \Eprint {http://arxiv.org/abs/arXiv:gr-qc/0810.1767}
  {arXiv:gr-qc/0810.1767} \BibitemShut {NoStop}%
\bibitem [{\citenamefont {{Szil{\'a}gyi}}(2014)}]{Szilagyi:2014fna}%
  \BibitemOpen
  \bibfield  {author} {\bibinfo {author} {\bibfnamefont {B.}~\bibnamefont
  {{Szil{\'a}gyi}}},\ }\href {\doibase 10.1142/S0218271814300146} {\bibfield
  {journal} {\bibinfo  {journal} {{Int. J. Mod. Phys. D}}\ }\textbf {\bibinfo
  {volume} {23}},\ \bibinfo {eid} {1430014} (\bibinfo {year} {2014})},\ \Eprint
  {http://arxiv.org/abs/1405.3693} {arXiv:1405.3693 [gr-qc]} \BibitemShut
  {NoStop}%
\bibitem [{\citenamefont {Boyle}\ and\ \citenamefont
  {Mrou{\'{e}}}(2009)}]{Boyle-Mroue:2008}%
  \BibitemOpen
  \bibfield  {author} {\bibinfo {author} {\bibfnamefont {M.}~\bibnamefont
  {Boyle}}\ and\ \bibinfo {author} {\bibfnamefont {A.~H.}\ \bibnamefont
  {Mrou{\'{e}}}},\ }\href {\doibase 10.1103/PhysRevD.80.124045} {\bibfield
  {journal} {\bibinfo  {journal} {Phys.\ Rev.\ D}\ }\textbf {\bibinfo {volume}
  {80}},\ \bibinfo {pages} {124045} (\bibinfo {year} {2009})},\ \Eprint
  {http://arxiv.org/abs/0905.3177} {arXiv:0905.3177 [gr-qc]} \BibitemShut
  {NoStop}%
\bibitem [{\citenamefont {Boyle}(2016)}]{Boyle2015a}%
  \BibitemOpen
  \bibfield  {author} {\bibinfo {author} {\bibfnamefont {M.}~\bibnamefont
  {Boyle}},\ }\href {\doibase 10.1103/PhysRevD.93.084031} {\bibfield  {journal}
  {\bibinfo  {journal} {Phys.\ Rev.\ D}\ }\textbf {\bibinfo {volume} {93}},\
  \bibinfo {pages} {084031} (\bibinfo {year} {2016})}\BibitemShut {NoStop}%
\bibitem [{\citenamefont {Boyle}(2013)}]{Boyle:2013a}%
  \BibitemOpen
  \bibfield  {author} {\bibinfo {author} {\bibfnamefont {M.}~\bibnamefont
  {Boyle}},\ }\href {\doibase 10.1103/PhysRevD.87.104006} {\bibfield  {journal}
  {\bibinfo  {journal} {Phys.\ Rev.\ D}\ }\textbf {\bibinfo {volume} {87}},\
  \bibinfo {pages} {104006} (\bibinfo {year} {2013})}\BibitemShut {NoStop}%
\bibitem [{\citenamefont {Boyle}\ \emph {et~al.}(2014)\citenamefont {Boyle},
  \citenamefont {Kidder}, \citenamefont {Ossokine},\ and\ \citenamefont
  {Pfeiffer}}]{Boyle:2014}%
  \BibitemOpen
  \bibfield  {author} {\bibinfo {author} {\bibfnamefont {M.}~\bibnamefont
  {Boyle}}, \bibinfo {author} {\bibfnamefont {L.~E.}\ \bibnamefont {Kidder}},
  \bibinfo {author} {\bibfnamefont {S.}~\bibnamefont {Ossokine}}, \ and\
  \bibinfo {author} {\bibfnamefont {H.~P.}\ \bibnamefont {Pfeiffer}},\
  }\href@noop {} {\  (\bibinfo {year} {2014})},\ \bibinfo {note}
  {arXiv:1409.4431},\ \Eprint {http://arxiv.org/abs/1409.4431}
  {arXiv:1409.4431} \BibitemShut {NoStop}%
\bibitem [{scr()}]{scri}%
  \BibitemOpen
  \href@noop {} {\enquote {\bibinfo {title} {Scri},}\ }\bibinfo {note}
  {\url{https://github.com/moble/scri}}\BibitemShut {NoStop}%
\bibitem [{\citenamefont {{Ossokine}}\ \emph {et~al.}(2015)\citenamefont
  {{Ossokine}}, \citenamefont {{Boyle}}, \citenamefont {{Kidder}},
  \citenamefont {{Pfeiffer}}, \citenamefont {{Scheel}},\ and\ \citenamefont
  {{Szil{\'a}gyi}}}]{Ossokine:2015vda}%
  \BibitemOpen
  \bibfield  {author} {\bibinfo {author} {\bibfnamefont {S.}~\bibnamefont
  {{Ossokine}}}, \bibinfo {author} {\bibfnamefont {M.}~\bibnamefont {{Boyle}}},
  \bibinfo {author} {\bibfnamefont {L.~E.}\ \bibnamefont {{Kidder}}}, \bibinfo
  {author} {\bibfnamefont {H.~P.}\ \bibnamefont {{Pfeiffer}}}, \bibinfo
  {author} {\bibfnamefont {M.~A.}\ \bibnamefont {{Scheel}}}, \ and\ \bibinfo
  {author} {\bibfnamefont {B.}~\bibnamefont {{Szil{\'a}gyi}}},\ }\href
  {\doibase 10.1103/PhysRevD.92.104028} {\bibfield  {journal} {\bibinfo
  {journal} {Phys.\ Rev.\ D}\ }\textbf {\bibinfo {volume} {92}},\ \bibinfo
  {eid} {104028} (\bibinfo {year} {2015})},\ \Eprint
  {http://arxiv.org/abs/1502.01747} {arXiv:1502.01747 [gr-qc]} \BibitemShut
  {NoStop}%
\bibitem [{\citenamefont {Boyle}\ \emph {et~al.}(2011)\citenamefont {Boyle},
  \citenamefont {Owen},\ and\ \citenamefont {Pfeiffer}}]{Boyle:2011gg}%
  \BibitemOpen
  \bibfield  {author} {\bibinfo {author} {\bibfnamefont {M.}~\bibnamefont
  {Boyle}}, \bibinfo {author} {\bibfnamefont {R.}~\bibnamefont {Owen}}, \ and\
  \bibinfo {author} {\bibfnamefont {H.~P.}\ \bibnamefont {Pfeiffer}},\ }\href
  {\doibase 10.1103/PhysRevD.84.124011} {\bibfield  {journal} {\bibinfo
  {journal} {Phys.\ Rev.\ D}\ }\textbf {\bibinfo {volume} {84}},\ \bibinfo
  {pages} {124011} (\bibinfo {year} {2011})},\ \Eprint
  {http://arxiv.org/abs/arXiv:1110.2965 [gr-qc]} {arXiv:1110.2965 [gr-qc]}
  \BibitemShut {NoStop}%
%%CITATION = ARXIV:1110.2965;%%
\bibitem [{\citenamefont {Aylott}\ \emph {et~al.}(2009)\citenamefont {Aylott},
  \citenamefont {Baker}, \citenamefont {Boggs}, \citenamefont {Boyle},
  \citenamefont {Brady} \emph {et~al.}}]{Aylott:2009ya}%
  \BibitemOpen
  \bibfield  {author} {\bibinfo {author} {\bibfnamefont {B.}~\bibnamefont
  {Aylott}}, \bibinfo {author} {\bibfnamefont {J.~G.}\ \bibnamefont {Baker}},
  \bibinfo {author} {\bibfnamefont {W.~D.}\ \bibnamefont {Boggs}}, \bibinfo
  {author} {\bibfnamefont {M.}~\bibnamefont {Boyle}}, \bibinfo {author}
  {\bibfnamefont {P.~R.}\ \bibnamefont {Brady}},  \emph {et~al.},\ }\href
  {\doibase 10.1088/0264-9381/26/16/165008} {\bibfield  {journal} {\bibinfo
  {journal} {Class.\ Quantum Grav.}\ }\textbf {\bibinfo {volume} {26}},\
  \bibinfo {pages} {165008} (\bibinfo {year} {2009})},\ \Eprint
  {http://arxiv.org/abs/0901.4399} {arXiv:0901.4399 [gr-qc]} \BibitemShut
  {NoStop}%
%%CITATION = ARXIV:0901.4399;%%
\bibitem [{\citenamefont {Smolyak}(1963)}]{smolyak1963quadrature}%
  \BibitemOpen
  \bibfield  {author} {\bibinfo {author} {\bibfnamefont {S.~A.}\ \bibnamefont
  {Smolyak}},\ }in\ \href@noop {} {\emph {\bibinfo {booktitle} {Dokl. Akad.
  Nauk SSSR}}},\ Vol.~\bibinfo {volume} {4}\ (\bibinfo {year} {1963})\ p.\
  \bibinfo {pages} {123}\BibitemShut {NoStop}%
\bibitem [{\citenamefont {Bungartz}\ and\ \citenamefont
  {Griebel}(2004)}]{bungartz2004sparse}%
  \BibitemOpen
  \bibfield  {author} {\bibinfo {author} {\bibfnamefont {H.-J.}\ \bibnamefont
  {Bungartz}}\ and\ \bibinfo {author} {\bibfnamefont {M.}~\bibnamefont
  {Griebel}},\ }\href@noop {} {\bibfield  {journal} {\bibinfo  {journal} {Acta
  numerica}\ }\textbf {\bibinfo {volume} {13}},\ \bibinfo {pages} {147}
  (\bibinfo {year} {2004})}\BibitemShut {NoStop}%
\bibitem [{\citenamefont {Schmidt}\ \emph {et~al.}(2011)\citenamefont
  {Schmidt}, \citenamefont {Hannam}, \citenamefont {Husa},\ and\ \citenamefont
  {Ajith}}]{Schmidt2010}%
  \BibitemOpen
  \bibfield  {author} {\bibinfo {author} {\bibfnamefont {P.}~\bibnamefont
  {Schmidt}}, \bibinfo {author} {\bibfnamefont {M.}~\bibnamefont {Hannam}},
  \bibinfo {author} {\bibfnamefont {S.}~\bibnamefont {Husa}}, \ and\ \bibinfo
  {author} {\bibfnamefont {P.}~\bibnamefont {Ajith}},\ }\href@noop {}
  {\bibfield  {journal} {\bibinfo  {journal} {Phys.\ Rev.\ D}\ }\textbf
  {\bibinfo {volume} {84}},\ \bibinfo {pages} {024046} (\bibinfo {year}
  {2011})},\ \Eprint {http://arxiv.org/abs/arxiv:1012.2879} {arxiv:1012.2879}
  \BibitemShut {NoStop}%
\bibitem [{\citenamefont {O'Shaughnessy}\ \emph {et~al.}(2011)\citenamefont
  {O'Shaughnessy}, \citenamefont {Vaishnav}, \citenamefont {Healy},
  \citenamefont {Meeks},\ and\ \citenamefont {Shoemaker}}]{OShaughnessy2011}%
  \BibitemOpen
  \bibfield  {author} {\bibinfo {author} {\bibfnamefont {R.}~\bibnamefont
  {O'Shaughnessy}}, \bibinfo {author} {\bibfnamefont {B.}~\bibnamefont
  {Vaishnav}}, \bibinfo {author} {\bibfnamefont {J.}~\bibnamefont {Healy}},
  \bibinfo {author} {\bibfnamefont {Z.}~\bibnamefont {Meeks}}, \ and\ \bibinfo
  {author} {\bibfnamefont {D.}~\bibnamefont {Shoemaker}},\ }\href {\doibase
  10.1103/PhysRevD.84.124002} {\bibfield  {journal} {\bibinfo  {journal} {Phys.
  Rev. D}\ }\textbf {\bibinfo {volume} {84}},\ \bibinfo {pages} {124002}
  (\bibinfo {year} {2011})},\ \Eprint {http://arxiv.org/abs/arXiv:1109.5224}
  {arXiv:1109.5224} \BibitemShut {NoStop}%
\bibitem [{\citenamefont {Blanchet}(2014)}]{Blanchet2014}%
  \BibitemOpen
  \bibfield  {author} {\bibinfo {author} {\bibfnamefont {L.}~\bibnamefont
  {Blanchet}},\ }\href {http://www.livingreviews.org/lrr-2014-2} {\bibfield
  {journal} {\bibinfo  {journal} {Living Rev.\ Rel.}\ }\textbf {\bibinfo
  {volume} {17}},\ \bibinfo {pages} {2} (\bibinfo {year} {2014})}\BibitemShut
  {NoStop}%
\bibitem [{\citenamefont {Butcher}(2003)}]{Butcher2003}%
  \BibitemOpen
  \bibfield  {author} {\bibinfo {author} {\bibfnamefont {J.~C.}\ \bibnamefont
  {Butcher}},\ }\href@noop {} {\emph {\bibinfo {title} {Numerical Methods for
  Ordinary Differential Equations}}}\ (\bibinfo  {publisher} {Wiley},\ \bibinfo
  {year} {2003})\BibitemShut {NoStop}%
\bibitem [{\citenamefont {Bashforth}\ and\ \citenamefont
  {Adams}(1883)}]{BA66330495}%
  \BibitemOpen
  \bibfield  {author} {\bibinfo {author} {\bibfnamefont {F.}~\bibnamefont
  {Bashforth}}\ and\ \bibinfo {author} {\bibfnamefont {J.~C.}\ \bibnamefont
  {Adams}},\ }\href {http://ci.nii.ac.jp/ncid/BA66330495} {\emph {\bibinfo
  {title} {An attempt to test the theories of capillary action by comparing the
  theoretical and measured forms of drops of fluid}}}\ (\bibinfo  {publisher}
  {Cambridge University Press},\ \bibinfo {year} {1883})\BibitemShut {NoStop}%
\bibitem [{SpE({\natexlab{b}})}]{SpECSurrogates}%
  \BibitemOpen
  \href@noop {} {}\bibinfo {howpublished}
  {\url{http://www.black-holes.org/surrogates/}} ({\natexlab{b}})\BibitemShut
  {NoStop}%
\bibitem [{\citenamefont {Babak}\ \emph {et~al.}(2017)\citenamefont {Babak},
  \citenamefont {Taracchini},\ and\ \citenamefont {Buonanno}}]{Babak:2016tgq}%
  \BibitemOpen
  \bibfield  {author} {\bibinfo {author} {\bibfnamefont {S.}~\bibnamefont
  {Babak}}, \bibinfo {author} {\bibfnamefont {A.}~\bibnamefont {Taracchini}}, \
  and\ \bibinfo {author} {\bibfnamefont {A.}~\bibnamefont {Buonanno}},\ }\href
  {\doibase 10.1103/PhysRevD.95.024010} {\bibfield  {journal} {\bibinfo
  {journal} {Phys. Rev.}\ }\textbf {\bibinfo {volume} {D95}},\ \bibinfo {pages}
  {024010} (\bibinfo {year} {2017})},\ \Eprint
  {http://arxiv.org/abs/1607.05661} {arXiv:1607.05661 [gr-qc]} \BibitemShut
  {NoStop}%
%%CITATION = ARXIV:1607.05661;%%
\bibitem [{\citenamefont {Shoemaker}(2010)}]{Shoemaker2009}%
  \BibitemOpen
  \bibfield  {author} {\bibinfo {author} {\bibfnamefont {D.}~\bibnamefont
  {Shoemaker}} (\bibinfo {collaboration} {{LIGO} Collaboration}),\ }\href
  {https://dcc.ligo.org/cgi-bin/DocDB/ShowDocument?docid=2974} {\enquote
  {\bibinfo {title} {Advanced {LIGO} anticipated sensitivity curves},}\ }
  (\bibinfo {year} {2010}),\ \bibinfo {note} {{LIGO} Document
  T0900288-v3}\BibitemShut {NoStop}%
\bibitem [{\citenamefont {Aasi}\ \emph {et~al.}(2015)\citenamefont {Aasi} \emph
  {et~al.}}]{aLIGO2}%
  \BibitemOpen
  \bibfield  {author} {\bibinfo {author} {\bibfnamefont {J.}~\bibnamefont
  {Aasi}} \emph {et~al.} (\bibinfo {collaboration} {LIGO Scientific
  Collaboration}),\ }\href {\doibase 10.1088/0264-9381/32/7/074001} {\bibfield
  {journal} {\bibinfo  {journal} {Class.\ Quantum Grav.}\ }\textbf {\bibinfo
  {volume} {32}},\ \bibinfo {pages} {074001} (\bibinfo {year} {2015})},\
  \Eprint {http://arxiv.org/abs/1411.4547} {arXiv:1411.4547 [gr-qc]}
  \BibitemShut {NoStop}%
%%CITATION = ARXIV:1411.4547;%%
\bibitem [{\citenamefont {Bustillo}\ \emph {et~al.}(2015)\citenamefont
  {Bustillo}, \citenamefont {Bohé}, \citenamefont {Husa}, \citenamefont
  {Sintes}, \citenamefont {Hannam} \emph {et~al.}}]{Bustillo:2015ova}%
  \BibitemOpen
  \bibfield  {author} {\bibinfo {author} {\bibfnamefont {J.~C.}\ \bibnamefont
  {Bustillo}}, \bibinfo {author} {\bibfnamefont {A.}~\bibnamefont {Bohé}},
  \bibinfo {author} {\bibfnamefont {S.}~\bibnamefont {Husa}}, \bibinfo {author}
  {\bibfnamefont {A.~M.}\ \bibnamefont {Sintes}}, \bibinfo {author}
  {\bibfnamefont {M.}~\bibnamefont {Hannam}},  \emph {et~al.},\ }\href@noop {}
  {\  (\bibinfo {year} {2015})},\ \Eprint {http://arxiv.org/abs/1501.00918}
  {arXiv:1501.00918 [gr-qc]} \BibitemShut {NoStop}%
%%CITATION = ARXIV:1501.00918;%%
\bibitem [{\citenamefont {Boyle}(2011)}]{Boyle:2011dy}%
  \BibitemOpen
  \bibfield  {author} {\bibinfo {author} {\bibfnamefont {M.}~\bibnamefont
  {Boyle}},\ }\href {\doibase 10.1103/PhysRevD.84.064013} {\bibfield  {journal}
  {\bibinfo  {journal} {Phys.\ Rev.\ D}\ }\textbf {\bibinfo {volume} {84}},\
  \bibinfo {pages} {064013} (\bibinfo {year} {2011})}\BibitemShut {NoStop}%
\bibitem [{\citenamefont {Ohme}\ \emph {et~al.}(2011)\citenamefont {Ohme},
  \citenamefont {Hannam},\ and\ \citenamefont {Husa}}]{OhmeEtAl:2011}%
  \BibitemOpen
  \bibfield  {author} {\bibinfo {author} {\bibfnamefont {F.}~\bibnamefont
  {Ohme}}, \bibinfo {author} {\bibfnamefont {M.}~\bibnamefont {Hannam}}, \ and\
  \bibinfo {author} {\bibfnamefont {S.}~\bibnamefont {Husa}},\ }\href {\doibase
  10.1103/PhysRevD.84.064029} {\bibfield  {journal} {\bibinfo  {journal}
  {Phys.\ Rev.\ D}\ }\textbf {\bibinfo {volume} {84}},\ \bibinfo {pages}
  {064029} (\bibinfo {year} {2011})}\BibitemShut {NoStop}%
\bibitem [{\citenamefont {{MacDonald}}\ \emph {et~al.}(2011)\citenamefont
  {{MacDonald}}, \citenamefont {Nissanke},\ and\ \citenamefont
  {Pfeiffer}}]{MacDonald:2011ne}%
  \BibitemOpen
  \bibfield  {author} {\bibinfo {author} {\bibfnamefont {I.}~\bibnamefont
  {{MacDonald}}}, \bibinfo {author} {\bibfnamefont {S.}~\bibnamefont
  {Nissanke}}, \ and\ \bibinfo {author} {\bibfnamefont {H.~P.}\ \bibnamefont
  {Pfeiffer}},\ }\href {\doibase 10.1088/0264-9381/28/13/134002} {\bibfield
  {journal} {\bibinfo  {journal} {Class.\ Quantum Grav.}\ }\textbf {\bibinfo
  {volume} {28}},\ \bibinfo {pages} {134002} (\bibinfo {year} {2011})},\
  \Eprint {http://arxiv.org/abs/1102.5128} {arXiv:1102.5128 [gr-qc]}
  \BibitemShut {NoStop}%
\end{thebibliography}%

\end{document}